\documentclass[useAMS,usenatbib]{mnras}

\usepackage[english]{babel}
\usepackage{amsmath,amssymb}
\usepackage{graphicx, subfig}
\usepackage[normalem]{ulem}

\usepackage{verbatim}

\usepackage{color}
\usepackage{multirow}
\usepackage{mathtools}
\usepackage{epstopdf}

\usepackage{url}
\usepackage{xcolor}

\usepackage{etoolbox}
\makeatletter
\newcount\c@additionalboxlevel
\newcount\c@maxboxlevel
\patchcmd\@combinedblfloats{\box\@outputbox}{%
  \stepcounter{additionalboxlevel}%
  \box\@outputbox
}{}{\errmessage{\noexpand\@combinedblfloats could not be patched}}

\AtBeginShipout{%
  \ifnum\value{additionalboxlevel}>\value{maxboxlevel}%
    \typeout{Warning: maxboxlevel might be too small, increase to %
      \the\value{additionalboxlevel}%
    }%
  \fi 
  \@whilenum\value{additionalboxlevel}<\value{maxboxlevel}\do{%
    \typeout{* Additional boxing of page `\thepage'}%
    \setbox\AtBeginShipoutBox=\hbox{\copy\AtBeginShipoutBox}%
    \stepcounter{additionalboxlevel}%
  }%
  \setcounter{additionalboxlevel}{0}%
}
\makeatother

\def\nat{Nature}
\def\na{New Astronomy}

\def\mnras{MNRAS}
\def\apj{ApJ}
\def\apjl{ApJL}
\def\apjs{ApJS}
\def\aap{A\&A}
\def\aaps{A\&A Supp.}
\def\aapr{A\&A Rev.}
\def\araa{ARA\&A}

\def\physrep{Phys. Rep.}
\def\pasa{Publ. Astr. Soc. Australia}

\def\ssr{Space Science Reviews}

\def\be{\begin{equation}} 
\def\ee{\end{equation}} 
\def\ba{\begin{eqnarray}} 
\def\ea{\end{eqnarray}}

\def\kms{\,{\rm {km\, s^{-1}}}} 
\def\cc{\,{\rm {cm^{-3}}}} 
\def\msun{{\Msun}}

\def\HH{${\rm {H_2}}$}

\def\HII{\hbox{H~$\scriptstyle\rm II\ $}}

\def\gsim{\lower.5ex\hbox{\gtsima}} 
\def\lsim{\lower.5ex\hbox{\ltsima}} \def\gtsima{$\; \buildrel > \over 
\sim \;$} \def\ltsima{$\; \buildrel < \over \sim \;$} \def\prosima{$\; 
\buildrel \propto \over \sim \;$} \def\gsim{\lower.5ex\hbox{\gtsima}} 
\def\lsim{\lower.5ex\hbox{\ltsima}} 
\def\simgt{\lower.5ex\hbox{\gtsima}} 
\def\simlt{\lower.5ex\hbox{\ltsima}} 
\def\simpr{\lower.5ex\hbox{\prosima}} \def\la{\lsim} \def\ga{\gsim} 
  
 \def\gtsima{$\; \buildrel > \over \sim \;$} 
\def\ltsima{$\; \buildrel < \over \sim \;$} 
\def\gsim{\lower.5ex\hbox{\gtsima}} 
\def\lsim{\lower.5ex\hbox{\ltsima}} 
\def\simgt{\lower.5ex\hbox{\gtsima}} 
\def\simlt{\lower.5ex\hbox{\ltsima}} 
\def\simpr{\lower.5ex\hbox{\prosima}} 
\def\la{\lsim} 
\def\ga{\gsim}

\def\msun{\,{\rm \Msun}}

\def\E3{{\cal E}_{\rm g}^{III}}

\def\r12{r_{1/2}} 
\def\x12{x_{1/2}} 
\def\v12{v_{1/2}}

\def\HII{\hbox{H~$\scriptstyle\rm II $~}}

\def\CIIion{${\rm C}^{+}$}
\def\OIIIion{${\rm O}^{++}$}
\def\NIIion{${\rm N}^{+}$}

\def\CII{\hbox{[C~$\scriptstyle\rm II $]}}
 
\def\OIII{\hbox{[O~$\scriptstyle\rm III $]}} 
\def\NII{\hbox{[N~$\scriptstyle\rm II $]}}

\newcommand\code[1]{\textsc{\MakeLowercase{#1}}}
\newcommand{\quotes}[1]{``#1''}

\def\nh2{n_{\rm H2}}
\def\fh2{f_{\rm H2}}

\def\Hp{${\rm H}^{+}$}
\def\dust{\mathcal{D}}
\def\mp{m_{\rm p}}

\def\angstrom{\textrm{A\kern -1.3ex\raisebox{0.6ex}{$^\circ$}}}
\def\myr{\rm Myr}

\def\msun{{\rm M}_{\odot}}
\def\zsun{{\rm Z}_{\odot}}
\def\lsun{{\rm L}_{\odot}}
\def\dsun{\dust_{\odot}}
\def\cc{{\rm cm}^{-3}}
\def\colcm{{\rm cm}^{-2}}
\def\msunyr{\msun\,{\rm yr}^{-1}}
\def\surfd{\msun\,{\rm kpc}^{-2}}
\def\surfsfr{\msun\,{\rm yr}^{-1}\,{\rm kpc}^{-2}}

\def\surfl{\lsun\,{\rm kpc}^{-2}}
\def\kpc{{\rm kpc}}
\def\SFR{{\rm SFR}}

\def\SCII{\Sigma_{\rm [CII]}}
\def\SOIII{\Sigma_{\rm [OIII]}}
\def\SNII{\Sigma_{\rm [NII]}}
\def\SigmaCII{\Sigma_{\rm C+}}
\def\SigmaOIII{\Sigma_{\rm O++}}
\def\SigmaNII{\Sigma_{\rm N+}}
\def\sigmasfr{\dot{\Sigma}_{\star}}

\def\highz{$\mbox{high-}z$~}

\makeatletter
\def\@hex@@Hex#1%
 {\if a#1A\else \if b#1B\else \if c#1C\else \if d#1D\else
  \if e#1E\else \if f#1F\else #1\fi\fi\fi\fi\fi\fi \@hex@Hex}
\makeatother

\definecolor{apcolor}{HTML}{b3003b}
\definecolor{cbcolor}{HTML}{ff0f00}
\definecolor{afcolor}{HTML}{b3443c}
\definecolor{ddcolor}{HTML}{077a2f}

\graphicspath{{plots_png/},{plots_png/map_overview/},{plots_png/eos/},{plots_png/map_emission/},{plots_png/comparison_obs/}}

\def\commento{}

\begin{document}

\date{}
\pagerange{\pageref{firstpage}--\pageref{lastpage}} \pubyear{2019}
\title[Structure of first galaxies]{Deep into the structure of the first galaxies: SERRA views}
\author[Pallottini et al.]{A. Pallottini$^{1,2}$\thanks{\href{mailto:andrea.pallottini@centrofermi.it}{andrea.pallottini@centrofermi.it}; \href{mailto:andrea.pallottini@sns.it}{andrea.pallottini@sns.it}},
A. Ferrara$^{2,3}$,
D. Decataldo$^{2}$,
S. Gallerani$^{2}$,
L. Vallini$^{4,5}$,
S. Carniani$^{2}$,\newauthor
C. Behrens$^{6}$,
M. Kohandel$^{2}$,
S. Salvadori$^{7,8}$.\\
$^{1}$ Centro Fermi, Museo Storico della Fisica e Centro Studi e Ricerche ``Enrico Fermi'', Piazza del Viminale 1, Roma, 00184, Italy\\
$^{2}$ Scuola Normale Superiore, Piazza dei Cavalieri 7, I-56126 Pisa, Italy\\
$^{3}$ Kavli Institute for the Physics and Mathematics of the Universe (WPI), University of Tokyo, Kashiwa 277-8583, Japan\\
$^{4}$ Leiden Observatory, Leiden University, PO Box 9500, 2300 RA Leiden, The Netherlands\\
$^{5}$ Nordita, KTH Royal Institute of Technology and Stockholm University Roslagstullsbacken 23, SE-106 91 Stockholm, Sweden\\
$^{6}$ Institut f\"{u}r Astrophysik, Georg-August Universit\"{a}t G\"{o}ttingen, Friedrich-Hundt-Platz 1, 37077, G\"{o}ttingen, Germany\\
$^{7}$ Dipartimento di Fisica e Astronomia, Universit\'{a} di Firenze, Via G. Sansone 1, Sesto Fiorentino, Italy\\
$^{8}$ INAF/Osservatorio Astrofisico di Arcetri, Largo E. Fermi 5, Firenze, Italy
}

\maketitle

\label{firstpage}

\begin{abstract}
We study the formation and evolution of a sample of Lyman Break Galaxies in the Epoch of Reionisation by using high-resolution ($\sim 10$ pc), cosmological zoom-in simulations part of the \code{SERRA} suite. In \code{SERRA}, we follow the interstellar medium (ISM) thermo-chemical non-equilibrium evolution, and perform on-the-fly radiative transfer of the interstellar radiation field (ISRF). The simulation outputs are post-processed to compute the emission of far infrared lines (\CII, \NII, and \OIII).
At $z= 8$, the most massive galaxy, \quotes{Freesia}, has an age $t_\star \simeq 409\,\myr$, stellar mass $M_{\star} \simeq 4.2\times 10^9 \msun$, and a star formation rate $\SFR \simeq 11.5\,\msunyr$, due to a recent burst. Freesia has two stellar components (A and B) separated by $\simeq 2.5\, \kpc$; other 11 galaxies are found within $56.9 \pm 21.6$ kpc. The mean ISRF in the Habing band is $G = 7.9\, G_0$ and is spatially uniform; in contrast, the ionisation parameter is $U = 2^{+20}_{-2} \times 10^{-3}$, and has a patchy distribution peaked at the location of star-forming sites. The resulting ionising escape fraction from Freesia is $f_{\rm esc}\simeq 2\%$. While \CII~emission is extended (radius 1.54 kpc), \OIII~is concentrated in Freesia-A (0.85 kpc), where the ratio $\SOIII/\SCII \simeq 10$.
As many high-$z$ galaxies, Freesia lies below the local \CII-SFR relation. We show that this is the general consequence of a starburst phase (pushing the galaxy above the Kennicutt-Schmidt relation) which disrupts/photodissociates the emitting molecular clouds around star-forming sites. Metallicity has a sub-dominant impact on the amplitude of \CII-SFR deviations.
\end{abstract}


\section{Introduction}

Characterising the interstellar medium (ISM) properties of galaxies in the epoch of the reionisation (EoR) represents a key quest of modern cosmology.

Optical/near infrared (IR) surveys have been fundamental in identifying galaxies in the EoR, and further to give us an overview of their stellar masses, star formation rates and sizes up to redshift $z \sim 10$, well within the EoR \citep{Dunlop13,Madau14,Bouwens:2015,oesch:2018}. In particular, lensing has enabled us to probe the faintest galaxies \citep{smit:14,bowens:2017,vanzella:2018}, that are likely the main responsible for the reionisation and metal enrichment of the intergalactic medium \citep{barkana:2001phr,ciardi:2005ssr,bromm:2011,pallottini:2014sim,greig:2017,dayal:2018,maiolino:2018rev}.

However, to understand the properties of the ISM of such objects, spectral information is needed. In particular, far infrared (FIR) lines can give a wealth of diagnostics on the thermo-dynamical state of the gas and on the interstellar radiation field (ISRF). As these lines are the main coolants of the ISM \citep{dalgarno:1972,wolfire:2003apj}, they can be used to trace feedback processes responsible for the evolution of these systems. Additionally, since FIR lines are emitted by ions with low (\CIIion) and high (\OIIIion) ionisation potential, their simultaneous detection can constrain the intensity and shape of the IRSF. Finally, detection of CO rotational transitions would help us to constrain the physical properties of molecular clouds \citep{solomon:2005,carilli:2013ara&a}, and thus understanding the processes of star formation in galaxies at the EoR.

The advent of the Atacama Large Millimeter/Submillmeter Array (ALMA) has made it possible to access FIR lines from \quotes{normal} star forming galaxies (${\rm SFR}\lsim 100\msunyr$) in the EoR.
In particular \CII~at ~$158\mu$, being typically the strongest FIR line \citep{stacey:1991}, is now routinely observed at $z\gsim 6$ in both Lyman Alpha Emitters \citep[LAE,][]{pentericci:2016apj,bradac:2017,matthee:2017,carniani:2018himiko,harikane:2018} and Lyman Break Galaxies \citep[LBG,][]{maiolino:2015arxiv,willott:2015arxiv15,capak:2015arxiv,knudsen:2016arxiv,carniani:2018}.
Additionally, \CII~follow-up observations have enabled us to study the galaxy kinematics \citep{jones+17,smit:2018}, albeit such observations have not yet the level of maturity as those concentrating on intermediate ($z\lsim 3$) redshift \citep[e.g.][]{debreuck:2014,leung:2018}.
Low surface brightness gas outside the target galaxy is possibly a tracer of outflows: probing such material would help us to constrain the feedback mechanism driving the evolution of EoR galaxies. However, so far only statistical evidence of its presence is currently available \citep{gallerani:2016outflow,fujimoto:2019}.
The presence of \OIII~at $88\mu$ has been revealed in various observations \citep{inoue:2016sci,laporte:2017apj,hashimoto:2018,tamura:2018}; in a few cases both \OIII~and \CII~have been simultaneously detected \citep{carniani:2017bdf3299,hashimoto:2018}, thus hindering the possibility to constrain the ISRF.
Regarding the detection of molecular lines, so far CO has been observed only in one normal star forming galaxy via a serendipitous detection \citep{dodorico:2018,feruglio:2018}.
Summarising, while a large progress has been made with respect to the first ALMA observation cycles, we currently do not have a complete picture of the FIR properties of these galaxies. It is still unclear whether the local relation between \CII~and star formation rate \citep[][]{delooze:2014aa,herreracamus:2015} is fulfilled by high redshift galaxies and which are the physical mechanisms responsible for its larger dispersion w.r.t. the local one \citep{carniani:2018}. Finally, a convincing explanation of the spectral shifts and spatial offsets between different lines and/or UV continuum that are often observed in these objects is still missing \citep{capak:2015arxiv,carniani:2017bdf3299}.

To address such issues, on the theoretical side, models of FIR emission from galaxies in the EoR have been developed; these models account for the typically lower metallicity of these systems, higher gas turbulence, and include the suppression of FIR emission by the CMB in low density gas \citep{vallini:2013mnras,olsen:2015apj,vallini:2015,vallini:2018,popping:2019}.
Such models account for the observed ISM and ISRF properties of these objects by post-processing numerical hydrodynamical simulations aimed at describing the formation and evolution of high-redshift galaxies \citep[see][for an extended discussion]{olsen:2018conf}.

Cosmological simulations -- and in particular zoom-in simulations -- have been used in order to study such galaxies. Most works concentrate on the relative importance of different kinds of feedback (e.g. SN, winds from massive stars, radiation pressure) in shaping early galaxy evolution \citep{agertz:2015apj,pallottini:2017dahlia,hopkins:2017}, the chemical evolution of these primeval systems \citep{maio:2016mnras,smith:2017mnras,pallottini:2017althaea,lupi:2018,capelo:2018}, the effect of radiation from local sources, the ISM ionisation state, and the consequences for the reionisation process \citep{katz:2016arxiv,trebitsch:2017,rosdahl:2018,hopkins:2018}.

In the past few years we have developed \code{SERRA}, a set of zoom-in simulations of LBGs in the EoR. Starting with \citet{pallottini:2017dahlia}, we zoomed-in on the structure of high-z galaxies by studying the formation of few galactic systems and following their evolution down to tens of parsec scales. Then, in \citet{pallottini:2017althaea}, we have analysed the impact of chemistry on the ISM by including thermo-chemical networks to follow the formation of \HH, that ultimately led to the formation of stars. Complementing this numerical simulations with both line \citep[][\CII, CO, Ly$\alpha$]{vallini:2015,vallini:2018,behrens:2019} and continuum \citep[][UV, IR]{behrens:2018} emission, we have been able to fairly compare our models with high-redshift observations. However, previous simulation were were lacking a consistent modelling of photoevaporation effects due to the ISRF, that can affect the emission properties of the FIR lines \citep{vallini:2017} and the evolution of molecular clouds \citep{decataldo:2017}.

With the aim of further improving our models, in the present work we include on-the-fly radiative transfer in our hydrodynamical simulations. By also including all the main sources of feedback (radiative, mechanical, chemical), we are able to pinpoint the origin of the deviation from the \CII-SFR relation that is observed for galaxies in the EoR.
Our numerical model is presented in Sec. \ref{sec_numerical}. An overview of the physical properties of our galaxy sample is given in Sec. \ref{sec_overview}. The FIR emission properties (\CII,~\OIII,~\NII) are covered in Sec. \ref{sec_emission_prop}, while Sec. \ref{sec_obs_comparison} focuses on the \CII-SFR relation. Conclusions are given in Sec. \ref{sec_conclusione}.

\section{Numerical simulations}\label{sec_numerical}

Our simulation suite \code{Serra}\footnote{Greenhouse in Italian.} is focused on zooming-in on galaxies in the EoR. In this work, we present \quotes{Freesia}, a prototypical LBG galaxy that is hosted by a $M_{\rm h}\simeq 10^{11} \msun$ dark matter (DM) halo at $z=6$. With respect to previous works \citep[][]{pallottini:2017dahlia,pallottini:2017althaea}, here we explore the effect of local sources of radiation.

Gas and DM evolution is simulated with a customised version of the Adaptive Mesh Refinement (AMR) code \code{ramses}\footnote{\url{https://bitbucket.org/rteyssie/ramses}\label{foot_ramses}} \citep{teyssier:2002}. In \code{ramses}, gas is tracked with a second-order Godunov scheme and particles evolution is computed with a particle-mesh solver \citep[see also][for the gravity solver]{guillet:2011Jcoph}.
Radiation coupling to hydrodynamics is performed with \code{ramses-rt} \citep{rosdahl:2013ramsesrt}, that solves photons advection within a momentum-based framework with the closure given by setting a M1 condition for the Eddington tensor \citep{aubert:2008}.
Coupling between gas and photons is handled by implementing a non-equilibrium chemical network generated with \code{krome}\footnote{\url{https://bitbucket.org/tgrassi/krome}\label{foot_krome}} \citep{grassi:2014mnras}.
Metal ion abundances and emission lines are calculated in post-processing, by interpolating grids of models obtained from the photo-ionisation code \code{cloudy} V17\footnote{\url{https://www.nublado.org}} \citep{ferland:2017}. The modelling for gas, radiation, stars and line emission is described in Sec.s \ref{sec_model_gas}, \ref{sec_model_stars}, \ref{sec_model_radiation}, and \ref{sec_emission} respectively.

\subsubsection*{Set-up}

We generate cosmological initial conditions (IC)\footnote{We assume cosmological parameters compatible with \emph{Planck} results: $\Lambda$CDM model with total matter, vacuum and baryonic densities in units of the critical density $\Omega_{\Lambda}= 0.692$, $\Omega_{m}= 0.308$, $\Omega_{b}= 0.0481$, Hubble constant $\rm H_0=100\, h\,{\rm km}\,{\rm s}^{-1}\,{\rm Mpc}^{-1}$ with $h=0.678$, spectral index $n=0.967$, $\sigma_{8}=0.826$ \citep[][]{planck:2013_xvi_parameters}.} at $z=100$ with \code{music} \citep{hahn:2011mnras}. The cosmological volume is $(20\,{\rm Mpc}/ h)^{3}$, and the base grid is resolved with a mass $m_b = 6\times 10^6\msun$ per gas resolution element. The Lagrangian volume of the target halo has a linear size of $2.1\,{\rm Mpc}/ h$ and is progressively refined by 3 concentric layers with increasing mass resolution, reaching a gas mass of $m_b = 1.2\times 10^4 \msun$ around the target halo. In this zoom-in region, we allow for 6 additional level of refinement by adopting a Lagrangian-like criterion. This enables us to reach scales of $l_{\rm res}\simeq 30\,{\rm pc}$ at $z=6$ in the densest regions, i.e. the most refined cells have mass and size typical of Galactic molecular clouds \citep[MC, e.g.][]{federrath:2013}. Note that the resolution and IC are the same used in \citet[][]{pallottini:2017dahlia,pallottini:2017althaea} to allow a fair comparison.

\subsection{Hydrodynamics}\label{sec_model_gas}

\subsubsection*{Chemical network}

As in \citet{pallottini:2017althaea}, we implement a non-equilibrium chemical network by using \code{krome} \citep{grassi:2014mnras}. The selected network includes H, H$^{+}$, H$^{-}$, He, He$^{+}$, He$^{++}$, H$_{2}$, H$_{2}^{+}$ and electrons. The network follows a total of 48 reactions\footnote{The reactions, their rates, and corresponding references are listed in App. B of \citep{bovino:2016aa}: we use reactions from 1 to 31 and 53, 54, from 58 to 61, and from P1 to P9; the rates are reported in Tab. B.1, Tab. B.2, and Tab. 2 of \citet{bovino:2016aa}, respectively.}, including photo-chemistry, dust processes and cosmic ray-induced reactions \citep[see also][for the original implementation]{bovino:2016aa}. Individual ICs for the various species and ions are computed accounting for the chemistry in a primordial Universe \citep{galli:1998AA}.

\subsubsection*{Metals and dust}

Metallicity ($Z$) is tracked as the sum of heavy elements, and we assume solar abundance ratios of the different metal species \citep{asplund:2009ara&a}.
Dust evolution is not explicitly tracked during simulation. We make the assumption that the dust-to-gas mass ratio scales with metallicity, i.e. $\dust = \dsun (Z/\zsun)$, where $\dsun/\zsun = 0.3$ for the Milky Way (MW) \citep[e.g.][]{hirashita:2002mnras}. While in principle it is possible to incorporate the evolution of dust grains in galaxy formation simulations \citep[e.g.][{\commento see also \citealt{asano:2013,derossi:2017} for semi-analytic models}]{grassi:2017dust,mcKinnon:2018}, this would bias the following comparison with \citet{pallottini:2017althaea} and will be explored in the future. The grain size distribution is important when modelling light extinction, and it is detailed in Sec. \ref{sec_model_radiation} (see in particular Fig. \ref{fig_dust_abs}).

Dust provides a formation channel for molecular hydrogen: the formation rate of \HH~on dust grains is approximated following \citet{Jura:1975apj}:
\be\label{eq_jura}
R_{\rm H2-dust} = 3\times 10^{-17}n\,n_{\rm H} (\dust/\dsun)\,\cc\,{\rm s}^{-1}\,,
\ee
where $n$ and $n_{\rm H}$ are the total and Hydrogen gas densities, respectively. We note that for $\dust\gsim 10^{-2}\dsun$ the dust channel is dominant with respect to gas-phase formation.

We adopt an initial metallicity floor $Z_{\rm floor}=10^{-3}\zsun$ since at $z \gsim 40$ our resolution does not allow us to reach a density high enough for efficient \HH~formation in the pristine gas of mini-halos, and consequently recover the formation of first stars \citep[e.g.][]{oshea:2015apj,smith_b:2018}. Such floor only marginally affects the gas cooling time and it is compatible with the metallicity of diffuse enriched IGM in cosmological metal enrichment simulations \citep[e.g.][]{pallottini:2014sim,maio:2015,jaacks:2018}.

To summarise, metals and dust are treated as passive scalars and we allow for metal enrichment by supernova (SN) explosions and by winds from massive stars (see Sec. \ref{sec_model_stars}).

\subsubsection*{Gas thermo-dynamics}

We model both the evolution of thermal and turbulent energy content of the gas.

The thermal energy is evolved by the thermo-chemical framework set with \code{krome} \citep[see][for details]{pallottini:2017althaea}. Note that photo-chemical reaction rates in each gas cell depend on the local amount of radiation and its energy distribution (see Sec. \ref{sec_model_radiation}).
Since metal species are not followed individually, we use the equilibrium metal line cooling function calculated via \code{cloudy} \citep{cloudy:2013} with a \citet{Haardt:2012} UV background.
Following cooling from individual metal species can change the thermodynamics of the low density ISM, but does not appreciably affect the star forming regions, as shown in \citet[][see also \citealt{gnedin:2012}]{capelo:2018}. While such change in the thermodynamical state of the gas can be important to correctly compute emission lines, we recall that in the present work this is accounted for in post-processing (Sec. \ref{sec_emission}).
Dust cooling is not explicitly included, however we note that it gives only a minor contribution to the gas temperature for $n<10^4\cc$ \citep[e.g.][in particular see their Fig.~3]{bovino:2016aa}.
We consider the contribution of cosmic microwave background (CMB), that effectively sets a temperature floor for the gas.

We model the turbulent energy content of the gas similarly to \citet{agertz:2015apj} \citep[see also][]{pallottini:2017dahlia}: turbulent (or non-thermal) energy density $e_{\rm nth}$ is injected in the gas by SN, winds and radiation pressure, and it is dissipated as \citep[][see eq. 2]{teyssier:2013mnras}
\be
  \dot{e}_{\rm nth} = - \frac{e_{\rm nth}}{ t_{\rm diss}}\,,
\ee
where $t_{\rm diss}$ is the dissipation time scale, which can be written as in \citet{maclow1999turb}
\be
  t_{\rm diss} = 9.785 \left( \frac{l_{\rm cell}}{100\,{\rm pc}}\right)\left(\frac{\sigma_{\rm turb}}{10\,{\rm km}\,{\rm s}^{-1}}\right) ^{-1}\,\myr\,,
\ee
where $\sigma_{\rm turb} = \sqrt{e_{\rm nth}}$ is the turbulent velocity dispersion.
Note that we do not explicitly include a source term due to shear \citep[see][for more refined turbulence models]{maier:2009,scannapieco:2010,iapichino:2017}.

\subsection{Radiation}\label{sec_model_radiation}

\begin{figure}
\centering
\includegraphics[width=0.485\textwidth]{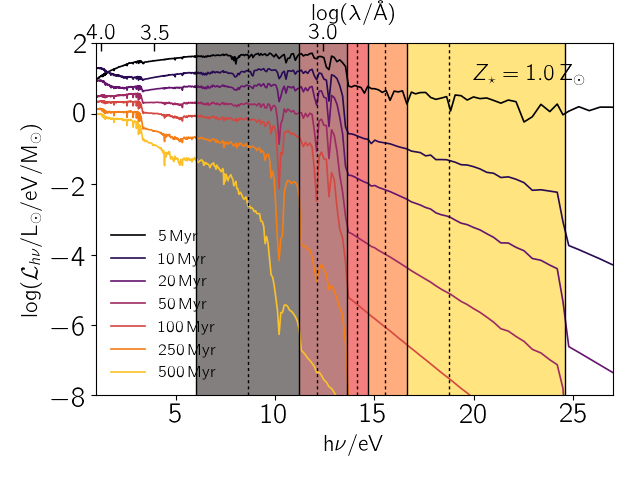}
\caption{
Stellar energy distribution (SED) per stellar mass ($\mathcal{L}_{\rm h \nu}$) per unit energy (${\rm h}\nu$) as a function of ${\rm h}\nu$ per a stellar population with $Z_{\star} = \zsun$. SEDs of different ages are plotted with different colours.
With shaded regions we highlight the photon energy bins considered in this work; in each bin dashed vertical lines indicates the photon energy averaged by weighting on a $t_\star = 10\,\myr$, $Z_{\star} = \zsun$ SED, i.e. the one assumed in the simulation to pre-calculate photon average quantities.
The wavelength ($\lambda$) corresponding to ${\rm h}\nu$ is indicated in the upper axis.
\label{fig_sed}
}
\end{figure}

\begin{figure}
\centering
\includegraphics[width=0.485\textwidth]{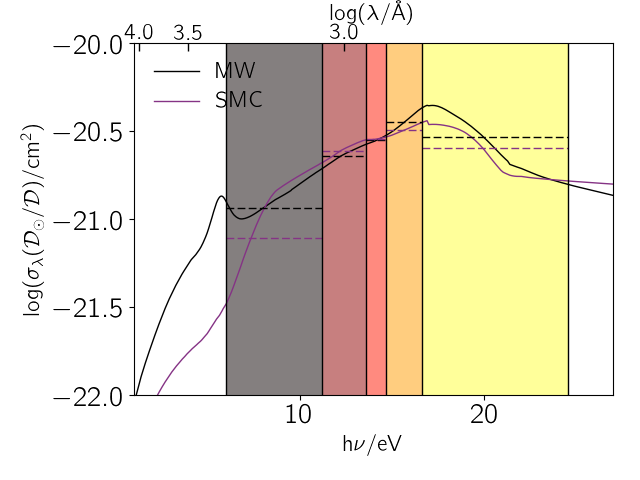}
\caption{
Dust cross section ($\sigma_\lambda$) per dust mass ($\dust$) in solar units ($\dsun$) as a function of photon energy (${\rm h}\nu$). Data from \citet{weingartner:2001apj} is plotted for a Milky Way and Small Magellanic Cloud -like composition. With shaded regions we highlight the photon energy bins considered in this work, and with dashed lines we indicate the SED averaged values in the bin.
The wavelength ($\lambda$) corresponding to ${\rm h}\nu$ is indicated in the upper axis.
\label{fig_dust_abs}
}
\end{figure}

In \code{ramses-rt}, photons are treated as a fluid that is spatially tracked by sharing the same AMR structure of the gas. Photons are separated in different energy bins, each one tracking an independent \quotes{fluid}.

For the present work we select 5 photon bins to cover both the energy range of the Galactic UV ISRF \citep{black:1987,draine:1978apjs} adopted in \citet{pallottini:2017althaea} and H ionising radiation\footnote{Photo-ionisation of He and He$^+$ is not included, as the stellar SED are typically not hard enough to produce such photons; He and He+ ionisation is only due to collision in the simulation.}. Fig. \ref{fig_sed} shows the stellar energy distribution (SED) per unit mass and unit energy for a stellar population at different age (see Sec. \ref{sec_model_stars} for details on the assumptions).
The first two low energy bins cover the \citet[][]{habing:1968} band\footnote{In this paper, the Habing flux $G$ is indicated in unit of $G_{0}=1.6\times 10^{-3} {\rm erg}\,{\rm cm}^{-2}\,{\rm s}^{-1}$, the MW value.} $6.0-13.6\,\rm eV$, which is fundamental in regulating the temperature of the ISM and photo-dissociation regions (PDR).
The second bin included in the Habing band is specific for the Lyman-Werner radiation ($11.2-13.6\,\rm eV$), which photo-dissociates \HH~via the two-step Solomon process \citep{stecher:1967apj}.
The last three bins cover the H-ionizing photons up to the first ionisation level of He ($13.6-24.59\,\rm eV$). For H-ionizing photons, the energy width is chosen such that the bins have the same number of photons when the SED is averaged on a fiducial stellar population. Our fiducial stellar population has an age $10\,\myr$ and $Z_{\star} = \zsun$: these stars are the main sources of the ISRF, since such young stars dominate the spectrum of a galaxy with an exponentially rising SFR, as expected for Freesia \citep[][]{pallottini:2017dahlia,pallottini:2017althaea}. To calculate the SED-averaged quantities, we fix the SED to the fiducial one.
Note that using three energy bins for H ionisation allows us to reasonably capture the temperature evolution of \Hp~regions, since photo-ionisation coupling with the gas is computed using the mean energy within each bin.

In \code{ramses} the time-step is determined by the Courant condition \citep{courant:1928}, i.e. $\delta t \sim l_{\rm cell}/v_g$, with $v_g$ being the gas velocity; taking $v_g$ of the order of the rotational velocity of the galaxy, this yields $\delta t \sim 30\, {\rm pc}/ 100 \, {\rm km}\,{\rm s}^{-1}\sim 0.3\,\myr$.
\code{ramses-rt} adopts an explicit scheme for the time evolution of radiation. Since hydrodynamics and radiation are coupled, it follows that the time-step of the simulation is determined by the minimum between the sound and light crossing time. To limit the computational load, we consider the reduced speed of light approximation to propagate wave-fronts \citep{gnedin:2001}, adopting $c_{red} = 10^{-2} c$. With such prescription, we expect $\delta t \sim l_{\rm cell}/c_{red}\sim 0.01\, \myr$.
Using a reduced speed of light approach yields artefacts when propagating light fronts very far from the sources, e.g. in IGM reionisation studies\footnote{Note that a more correct treatment on large scales is possible with a variable speed of light approach \citep{katz:2016arxiv}.}; however, it well captures the radiation transfer in the ISM/CGM of galaxies \citep[see][for a detailed study of the impact of a reduced speed of light]{deparis:2018}, as such it is well suited for the present work.

\subsubsection*{Coupling with gas and dust}

In the original implementation of \code{ramses-rt}, the thermo-chemical time step is performed simultaneously with the radiation propagation and absorption of photons by gas and dust. Such coupled step is sub-cycled in order to ensure simultaneous convergence for the absorbed photons, final ionisation state, and gas temperatures. This scheme is similar to the one adopted by \citet{nickerson:2018}, which includes \HH~formation in \code{ramses-rt}, albeit ${\rm H}^{-}$ and ${\rm H}_{2}^{+}$ are not explicitly followed.

Here we split the convergence steps: similarly to \code{ramses-rt} the absorption of photons is sub-cycled; for each of these sub-cycles, we obtain the ionisation state and gas temperatures with \code{krome}, that adaptively solves the thermo-chemical time evolution assuming a constant impinging flux.

For the gas absorption, photon cross sections are the same ones used in the chemical network (see Sec. \ref{sec_model_gas}).
For dust we assume a MW-like grain composition from \citet{weingartner:2001apj}. Both for gas and dust, the cross sections $\sigma_{\lambda}$ are used to pre-compute the $i$-th cross section in the photon energy bin ${\rm h} \nu_i^{\rm low}-{\rm h} \nu_{i}^{\rm up}$ as
\be
  \sigma_{\rm i} = \frac{\int_{\nu_i^{\rm low}}^{\nu_{i}^{\rm up}} \sigma_\lambda L_{\rm h \nu} {\rm d}\nu}{\int_{\nu_i^{\rm low}}^{\nu_{i}^{\rm up}} L_{\rm h \nu} {\rm d}\nu }\,,
\ee
i.e. flux-averaged cross sections, with a weight given by the selected impinging SED $L_{\rm h \nu}$. In Fig. \ref{fig_dust_abs} we plot the dust absorption cross section ($\sigma_\lambda$) per unit of $\dust/\dsun$ for both the MW and Small Magellanic Cloud (SMC) -like dust composition, as a function of photon energy ($h\nu$). For both dust types, the $\sigma_{i}$ are overplotted with dashed lines. The difference in absorption between MW- and SMC-like is $\lsim 1\%$, except in the $6.0-11.2\,\rm eV$ band, where the difference is about $\simeq 30\%$.

{\commento The analysis of current data seems to favour a MW-like distribution for \highz~galaxies. \citet{behrens:2018} manage to explain the observation by \citet{laporte:2017apj} with a low amount of MW-like dust, that leads to a warm FIR SED. However, the situation is still unclear. \citet{derossi:2017} analyses the role of silicate rich dust, that is expected from a Pop III dominated enrichment: the work shows that a warmer FIR SED can naturally arise because of the emission properties of silicates, that typically emit at higher wavelength with respect to the carbonaceous grains. Further, in \citet{derossi:2018} it is shown that a combination of silicate and small amount of carbonaceous grains can reproduce the SED observed in Haro 11, a local low-metallicity starburst, thought to be an analogue of high-z galaxies: this entails that assuming a simplified dust models can possibly modify the properties inferred from high-z observations \citep[cfr.][]{behrens:2018}. While dust composition (silicate vs carbonaceous grains) does change the FIR emission properties, in the present work dust is considered only with respect to its absorption properties, that are mainly dependent on the grain size distribution which is assumed to be time-independent (see Sec. \ref{sec_model_gas}). The analysis of the possible modification to the FIR emission due to a different dust composition is left for a future work.

Summarizing, while a different assumption on the dust distribution can in principle heavily affect the observed SED, it should produce only minor differences in the adopted model, since} the only energy bin where $\sigma_i$ is different in the two cases is responsible for neither \HH~dissociation nor H~ionisation.
Finally, the self-shielding of \HH~from photo-dissociation is accounted for by using the \citet{richings:2014} prescription, given the \HH~column density, temperature and turbulence \citep[cfr. with][]{wolcottgreen:2011}, as detailed in \citet{pallottini:2017althaea}.

While our scheme is different with respect to the one presented in \citet{rosdahl:2013ramsesrt} and \citet{nickerson:2018}, the overall results are consistent; detailed tests of our adopted scheme are found in the Appendix of \citet{decataldo:2019}, using PDR \citep[][]{rollig:2007,nickerson:2018} and \HII region \citep{iliev:2009} benchmarks.

\subsection{Stars}\label{sec_model_stars}

\subsubsection*{Formation}

As in \citet{pallottini:2017althaea}, stars form according to a Kennicutt-Schmidt-like relation \citep[][]{schmidt:1959apj,kennicutt:1998apj} that depends on the molecular hydrogen density ($\nh2$):
\be\label{eq_sk_relation}
\dot{\rho}_{\star}= \zeta_{\rm sf} \frac{\mu \mp \nh2}{t_{\rm ff}},
\ee
where $\dot{\rho}_{\star}$ is the local star formation rate density, $\zeta_{\rm sf}$ the star formation efficiency, $\mp$ the proton mass, $\mu$ the mean molecular weight, and $t_{\rm ff}$ the free-fall time.
The efficiency is set to $\zeta_{\rm sf}=10\%$, by adopting the average value observed for MCs \citep[][see also \citealt{agertz:2012arxiv}]{murray:2011apj}, while $n_{\rm H2}$ computation is included in the non-equilibrium chemical network.
As done in \citet{rasera:2006,dubois:2008,pallottini:2014sim}, eq. \ref{eq_sk_relation} is solved stochastically at each time step $\delta t$ in each cell with size $\delta l$, by forming in each possible event a new star particle with mass $m_\star = m_b\,N_\star $, with $N_\star$ drawn from a Poisson distribution characterised by mean
\be\label{eq_mean_poisson_sfr}
\langle N_\star\rangle = \frac{\mu \mp \nh2 (\delta l)^3}{m_b} \frac{\zeta_{\rm sf} \delta t}{t_{\rm ff} }\,.
\ee
For numerical stability, no more than half of the cell mass is allowed to turn into a star particle. Additionally, we allow only star formation events that spawn stellar clusters with mass $m_\star\geq 1.2\times 10^4 \msun$, i.e. the gas mass resolution of the simulation.

\subsubsection*{Stellar populations}

A single star particle in our simulations can be considered a stellar cluster, with metallicity $Z_{\star}$ set equal to that of the parent cell. For the stellar cluster, we assume a \citet{kroupa:2001} initial mass function and, by using \code{starburst99} \citep{starburst99:1999}, we adopt single population stellar evolutionary tracks given by the {\tt padova} \citep{padova:1994} library, that covers the $0.02 \leq Z_{\star}/\zsun \leq 1$ metallicity range. The stellar tracks are then used to calculate mechanical, chemical, and radiative feedback.

\subsubsection*{Mechanical and chemical feedback}

As in \citet{pallottini:2017dahlia}, we account for stellar energy inputs and chemical yields that depend both on metallicity $Z_\star$ and age $t_\star$ of the stellar cluster. Stellar feedback includes SNe, winds from massive stars, and radiation pressure \citep[see also][]{agertz:2012arxiv}.

Depending on the kind of feedback, stellar energy input can be both thermal and kinetic, and we account for the dissipation of energy in MCs for SN blastwaves \citep{ostriker:1988rvmp} and OB/AGB stellar winds \citep{weaver:1977apj}, as detailed in Sec.~2.4 and Appendix A of \citet{pallottini:2017dahlia}.

In \citet{pallottini:2017dahlia} radiation pressure was implemented by adding a source term to the turbulent (non-thermal) energy. Thus, to avoid double counting of such feedback, we turn off the original radiation pressure coupling of \code{ramses-rt}, that is done by following an extra infrared energy bin \citep[see][]{rosdahl2015radpress}.

\subsubsection*{Radiative feedback}

Stellar tracks are used to calculate photon production. At each time step, stars act as a source, dumping photons in each energy bin according to their the stellar age and metallicity (Sec. \ref{sec_model_radiation}, in particular Fig. \ref{fig_sed}), then photons are advected and absorbed in the radiation step, contributing at the same time to the photo-chemistry.

We neglect the cosmic UVB, since the typical ISM densities are sufficiently large to ensure an efficient self-shielding \citep[e.g.][]{gnedin:2010}. For example, \citet{rahmati:2013mnras} have shown that at $z\simeq 5$ the hydrogen ionisation due to the UVB is negligible for $n \gsim 10^{-2}\cc$, the typical density of diffuse ISM.

We do not explicitly consider production of radiation from recombination, i.e. we assume that recombination photons are absorbed \quotes{on the spot}, which is a valid approximation in the optically thick regime \citep{rosdahl:2013ramsesrt}.

Cosmic-ray (CR) processes are not explicitly tracked during the simulation \citep[see however][for possible implementations]{dubois:2016,pfrommer2017}. Similarly to \citet{pallottini:2017althaea}, we assume a CR hydrogen ionisation rate proportional to the global SFR \citep{valle:2002apj} and normalised to the MW value \citep[][see \citealt{ivlev:2015} for the spectral dependence]{webber:1998apj}:
\be
\zeta_{\rm cr} = 3\times 10^{-17} ({\rm SFR}/\msunyr)\, {\rm s}^{-1}.
\ee
Coulomb heating is accounted for by assuming that every CR ionisation releases an energy of $20$ eV \citep[see][ for a more accurate treatment]{glassgold:2012apj}.

\subsection{Ions and emission lines}\label{sec_emission}
\begin{figure*}
\centering
\includegraphics[width=0.485\textwidth]{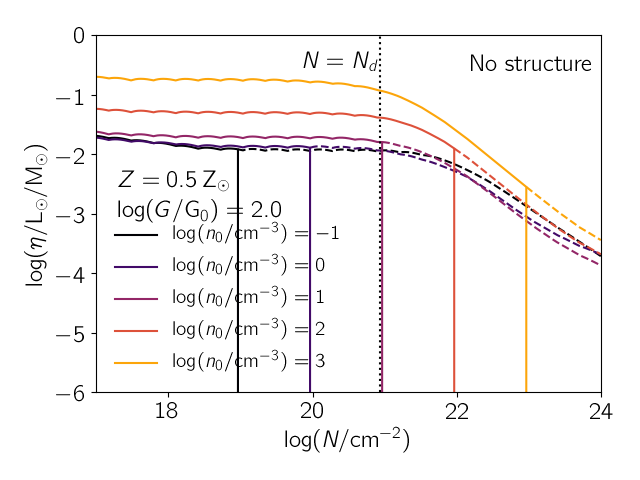}
\includegraphics[width=0.485\textwidth]{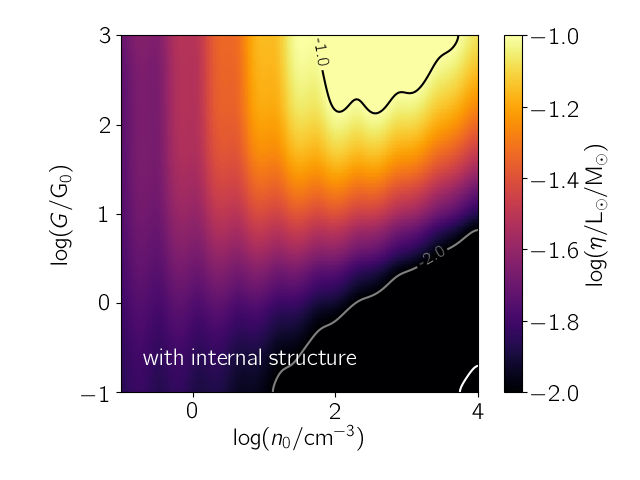}
\caption{
{\bf Left panel} Efficiency of \CII~emission ($\eta = L/M$) as a function of column density for clouds with fixed metallicity $Z=0.5\,\zsun$ and impinging non-ionizing ISRF $G=10^2 G_{0}$. The models have different density $n_0$ as indicated in the inset and are calculated with \code{cloudy}. For each model, the line turns from solid to dashed when reaching $N = n_0 \times 30\, {\rm pc}$, i.e. at the reference resolution of the simulation; this mark is also highlighted by vertical lines. The dotted vertical line indicates when $N=N_d$ (see eq. \ref{eq_dust_density})
{\bf Right panel} Efficiency of \CII~as a function of $G$ and $n_0$ when the internal structure is included (eqs. \ref{eqs_cloud_structure} and \ref{eqs_cloud_emission}). Models are computed assuming a Mach number $\mathcal{M}=10$, $Z=0.5\,\zsun$, and for $T=20\,\rm K$. The $\eta$ dynamical range has been restricted for visualisation purposes, and contours have been added to guide the eye.
\label{fig_benchmark_cloudy}
}
\end{figure*}

We model metal ion abundances and line emissions using \code{cloudy} V17 \citep{ferland:2017} in post-processing.
However, there are some typical challenges and shortcomings to consider when combining emission line codes with simulations \citep[see][for an overview]{olsen:2018conf}.

In the following we elaborate on the fact that i) a direct post-processing is computationally unfeasible and not completely consistent, ii) resolution limits our possibility to recover the ionising radiation and internal structure of molecular clouds. We conclude with the description of the solution adopted for the present work.

\subsubsection*{Consistency of the post-processing}

In the simulation, we solve non-equilibrium photo-chemistry by coupling \code{ramses-rt} and \code{krome}, while \code{cloudy} calculations assume photo-ionisation equilibrium. Moreover, \code{cloudy} does not account for dynamical effects -- such as e.g. shocks  -- which might affect ion abundances and emission line intensities \citep[cfr.][]{sutherland:2017,mappingsV:2018}. However, the typical PDR code \citep{rollig:2007} includes -- and models more accurately -- a larger number of physical processes with respect to the ones typically considered in hydrodynamical simulations with radiative transfer.

One option for the post-processing would consist in using the ISRF resulting from the simulation, and apply a single \code{cloudy} model to each cell, given its ISM physical characteristics.
On the one hand this is costly: a \code{cloudy} model is completed to convergence in $\sim 0.1-0.5$ CPU hours, depending on the chosen maximum column density. Since each snapshot typically contains $\sim 10^7 - 10^8$ cells, the cost in CPU hours to post-process a simulation snapshot would be comparable to the cost of the simulation itself \citep[see also][for a machine learning approach to the problem]{katz:2019arXiv}.
On the other hand, it is not guaranteed that such approach would result in a more consistent result. For instance in the hydrodynamical simulation we adopt 5 spectral energy bins, which is a very sparse sampling of the SED when compared to photo-ionisation code calculations, where typically thousands of energy bins are included.
Moreover, in a \code{cloudy} calculation, a cell is divided in optically thin slices, and the temperature and chemical structure is then calculated as a function of the optical depth. Also, \code{cloudy} integrates up to photo-ionisation equilibrium to resolve the internal structure of gas patches (single cells in the simulation). Thus, differences in ion abundances are expected with respect to the adopted \code{krome} scheme.

\subsubsection*{Limits given by the resolution}

The resolution and refinement criterion of the hydrodinamical simulation do no guaranteed to resolve dense~\Hp~regions. The column density of \Hp~in a slab of a dusty gas can be written as \citep{ferrara:2019}
\begin{subequations}\label{eqs_stima_ionization}
\be
N_{\rm H+} \simeq N_d \log(1+ 59 U\,{\cal D}/\dsun)\,,
\ee
where $U$ is the ionisation parameter and
\be\label{eq_dust_density}
N_d \simeq 1.7\times 10^{21} (\dsun/{\cal D})\,\colcm
\ee
\end{subequations}
is the column density due to dust where the optical depth to UV photons becomes unity (see Fig. \ref{fig_dust_abs}). From \citet{pallottini:2017althaea}, we expect the dense ISM of our galaxy to have $n = 3\times 10^{2}\cc$, $Z=0.5\,\zsun$, and $N \simeq 10^{22}\colcm$, because of our $\simeq 10\,\rm pc$ resolution. Using these values and eq.s \ref{eqs_stima_ionization}, in a typical photo-ionisation region ($U\simeq 10^{-2}$) we obtain a ionisation fraction $f_{\rm H+} \simeq 10\%$.
A partial ionisation in a single cell implies that all the ionising photons are absorbed, since in \code{ramses-rt} photons are advected after the absorption step\footnote{This is equivalent to state that \code{ramses-rt} keeps track of the absorbed flux and not the impinging one; the difference between is negligible only for optically thin cells.} Thus, it is possible to find young star clusters embedded in dense gas patches that have $f_{\rm H+}>0$ and $U=0$.

Moreover, with our $\simeq 10 \, {\rm pc}$ resolution we cannot resolve the internal structure of molecular clouds (MC), that are made of clumps and cores of size $\lsim 0.1 \rm pc$. Accounting for such contribution is important to correctly compute the emission in high density ($n\sim 10^2 \cc$) regions illuminated by a strong ($G\sim 10^3 G_{0}$) and ionising ($U\sim 10^{-2}$) ISRF \citep{vallini:2017}. These ISM regions are expected to be the main contributors of various FIR lines (i.e. \OIII) in \highz~galaxies \citep[e.g.][]{carniani:2017bdf3299} and their lower~$z$ analogues \citep[e.g.][]{cormier:2012aa}.

\subsubsection*{Adopted model}

To summarise, the limited resolution of a typical galaxy simulation does not allow us to recover the physical ion structure/line emission even if we would run a single \code{cloudy} model per cell, which 1) is computationally very expensive and 2) the assumptions are not completely consistent with the one adopted in the run. To overcome these problems we have adopted a different strategy, as described below.

Two distinct grids of models are calculated, with and without ionising radiation\footnote{In practical terms, the \code{cloudy} models without ionising radiation are calculated by interposing a dust-free obscuring screen with column density of $N_{H}=10^{23}\colcm$ between the source and the gas.}.
Parameters for each grid are $n$, $G$, and $Z$, with the following ranges: $10^{-1}\leq n/\cc \leq 10^{4.5}$, $10^{-1}\leq G/{\rm G}_{0} \leq 10^{4.5}$, $10^{-3}\leq Z/\zsun \leq 10^{0.5}$; for each parameter the grid spacing is $0.5$ dex, thus there are a total of 1152 individual models per grid.
As assumed in the simulation, dust is proportional to metallicity. We use an impinging SED taken from \code{starburst99} \citep{starburst99:1999} with age $t_{\star} = 10\,\myr$, metallicity $Z_{\star}=\zsun$, and rescaled with $G$.
Additionally to the ISRF, we include the CMB at the appropriate redshift. Note that \code{cloudy} V17 explicitly considers the CMB suppression \citep{dacunha:2013apj,vallini:2015,pallottini:2015cmb} and does subtract isotropic backgrounds, similarly to what is done in an ALMA observation \citep{ferland:2017}.
For each model we stop each calculation at $N=10^{23} \colcm$, after convergence has been reached.

Given $n$, $G$, $Z$, and $N$ in each cell, the ion abundances and emission lines can be interpolated from the values found in the computed grid.
The grid that includes ionising radiation is selected for those gas patches that either have a ionisation parameter $U>U_{\rm th}\equiv 10^{-4}$ or contain young ($t_{\star}<10\,\myr$) star clusters. The grid with non ionising radiation is chosen for all the other cells.
For lines arising from high-ionisation state (i.e. \OIII), this method allows us to recover both the emission from the diffuse ionised medium and from possibly unresolved dense ionised regions.
Changes in the selected $U_{\rm th}$ threshold do not yield large variations in the \OIII~total luminosity, since (i) low $U$ entails low flux, while high radiation fields are needed to produce a substantial emission \citep[$G\gsim 10^2 G_{0}$, see][]{vallini:2017}, (ii) regions containing young star clusters dominate the FIR emission of highly-ionised species \citep{cormier:2012aa}; this point is detailed in the results, i.e. Sec. \ref{sec_emission_prop} (in particular see Fig. \ref{fig_eos_emission}).

To account for the internal structure of MCs, we use a model similar to \citet{vallini:2015} \citep[see also][]{vallini:2017,vallini:2018}. We assume that a MC with mean density $n_0$ and mach number $\mathcal{M}$ is characterised by a probability density function (PDF) given by a log-normal distribution \citep{padoan:2011}:
\begin{subequations}\label{eqs_cloud_structure}
\be
{\rm d}P \equiv  P_s {\rm d}s = (2\pi \sigma_s^{2})^{-1/2} \exp^{-\left(\frac{s-s_0}{\sqrt{2}\sigma_s}\right)^2}\, ,
\ee
with $s$ being the logarithmic density
\be
s = \ln({n/n_0})\,,
\ee
and where the constants $s_{0}$ and $\sigma_s$ are given by
\be
s_{0} = - \sigma_s^2/2\,
\ee
and
\be
\sigma_s^2 = \ln (1 + (\mathcal{M}/2)^2)\,,
\ee
respectively \citep[see also][]{krumholz:2005,tasker:2009,molina:2012,federrath:2013}.
\end{subequations}

In each MC, we assume that individual clumps have size given by the Jeans length $l_J = l_J(n,T)$, and volume $l_J^3$.
Then, the differential number of clumps of a MC with total volume $V$ can be written as
\begin{subequations}\label{eqs_cloud_emission}
\be
d N_{cl} = (V /l_J^3) {\rm d}P \,.
\ee
With our grids of \code{cloudy} models we can compute $F_{k} = F_{k}(n,G,Z,N)$ and $x_{y}=f_y(n,G,Z,N)$, i.e. the luminosity per unit area for line $k$ and the ion mass fraction per unit volume of ion $y$ for each clump. Then the total luminosity and ion mass of the MC can be written as
\be
L_k = \int F_k l_J^2 dN_{cl} \,
\ee
and
\be
M_y = \int x_y l_J^2 dN_{cl} \,,
\ee
respectively.
\end{subequations}

\subsubsection*{Case example}

It is interesting to discuss a case example of such model for \CII~by looking at the efficiency $\eta = L_{\rm CII}/M$, i.e. the \CII~luminosity to total gas mass ratio of the MC.

We start by analysing single cloud models -- i.e. without internal structure -- for different $n_0$ with fixed metallicity ($Z=0.5 \,\zsun$) and a non-ionizing radiation field of intensity $G=10^{2}G_{0}$. The value of $\eta$ as a function of $N$ is plotted in the left panel of Fig. \ref{fig_benchmark_cloudy}. At fixed $n_0$ for $N<N_d$, roughly $F_{\rm CII}\propto N =n_0 l_0$, thus $\eta = L/M \propto F_{\rm CII} l_0^2/l_0^3$ is constant, with $l_0$ being the size of the cloud; for $N>N_d$, $\eta$ drops in all cases, because $F_{\rm CII}$ becomes constant, and thus $\eta \propto 1/l_0$.

An important corollary of this result is that, applying these arguments to a simulation with a fixed resolution $l_{\rm res}$ will likely underestimate \CII~emission for $N>N_d$, i.e. in cells with resolution (see eq. \ref{eq_dust_density})
\be\label{eq_res_required}
l_{\rm res} > N_d/n \simeq 5.5 \,( n / 100\,\cc)^{-1}\,(Z / \zsun)^{-1}\rm pc\,.
\ee
For typical current state-of-the-art galaxy simulations such value is very demanding, rarely reached, and not compatible with typical refinement criterion selected. Subgrid-models are needed to correctly recover and predict the emission arising from the internal structure of MCs.

To this aim, we use our model for the internal structure of MC (eq.s \ref{eqs_cloud_structure}) and recompute the expected \CII~emission (eq. \ref{eqs_cloud_emission}). The result is presented in the right panel of Fig. \ref{fig_benchmark_cloudy} as a function of $n$ and $G_{0}$ for $\mathcal{M}=10$ and $Z=0.5\,\zsun$, i.e. the typical figures found in our simulated galaxies \citep{pallottini:2017althaea,vallini:2018}. A maximum value, $\eta\simeq 10^{-1}\lsun/\msun$, fall approximately at $n\simeq 10^3\cc$ and $G\gsim 5\times 10^2 G_{0}$, while at high density ($n>10^2\cc$) and low radiation field ($G\lsim 10 G_{0}$) the \CII~emission is inefficient ($\eta\lsim 10^{-2}\lsun/\msun$).
Note that as at low densities ($n\lsim 10\,\cc$) MCs have little internal structure, the sub-grid model result coincide with single cloud ones.

Adopting such sub-grid model is not completely self-consistent, since \code{cloudy} uses an equilibrium approximation, different binning for ISRF, and more physical processes/chemical species. However, it heals some of the problems affecting the calculation of line emission from post-processing of galaxy simulations that cannot spatially resolve \Hp~regions and the internal structure of MCs.

\section{Overview of the results}\label{sec_overview}

\begin{figure*}
\includegraphics[width=0.98\textwidth]{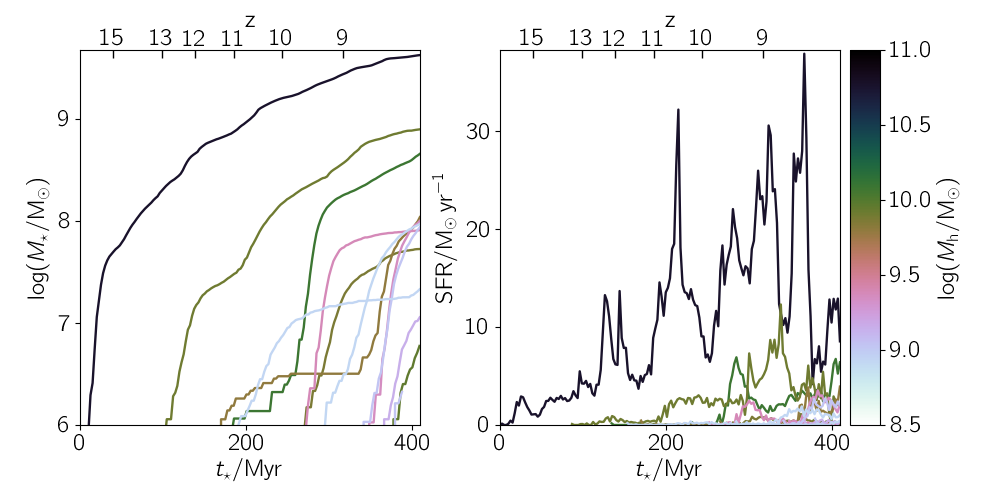}
\caption{Stellar mass build up ($M_{\star}$, {\bf left panel}) the star formation history (SFR, {\bf right panel}) as a function of the galaxy age ($t_\star$) and redshift ($z$, upper axis). Each simulated galaxy is plotted as an individual solid line colored accordingly to the hosting dark matter halo mass ($M_{\rm h}$); Freesia, the most massive galaxy in our sample is coloured in black. The reference for $t_\star=0$ is the first star formation event in Freesia ($z\sim 16$). Note that galaxies are defined as the collection of stellar particles belonging to the same dark matter halo at $z=8$; thus, part of the early star formation history shown can have took place in halo components at that were separated at that time.
\label{fig_time_evol_stars}
}
\end{figure*}

\begin{figure*}
\includegraphics[width=0.32\textwidth]{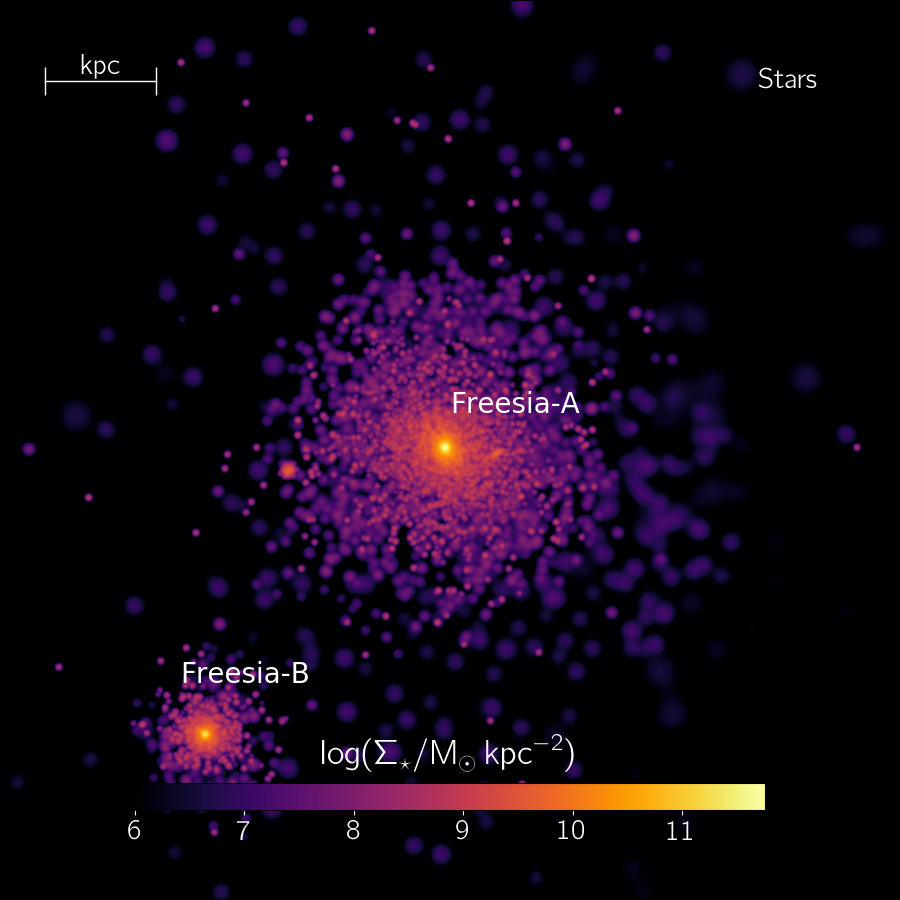}
\includegraphics[width=0.32\textwidth]{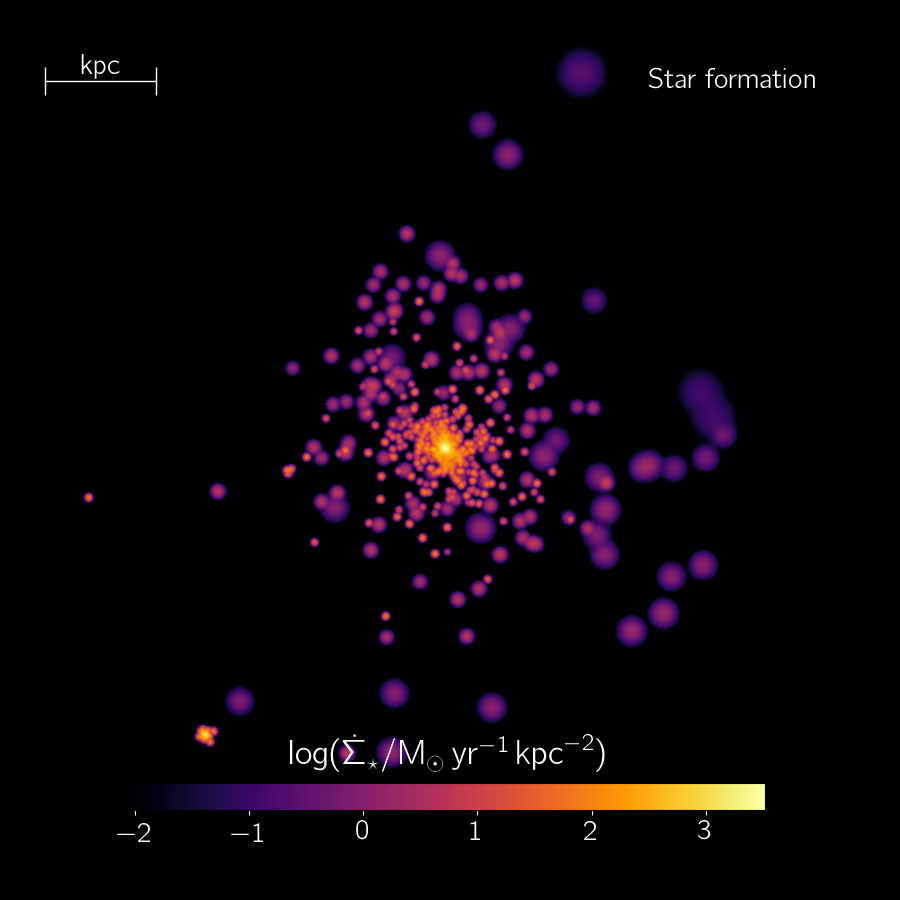}
\includegraphics[width=0.32\textwidth]{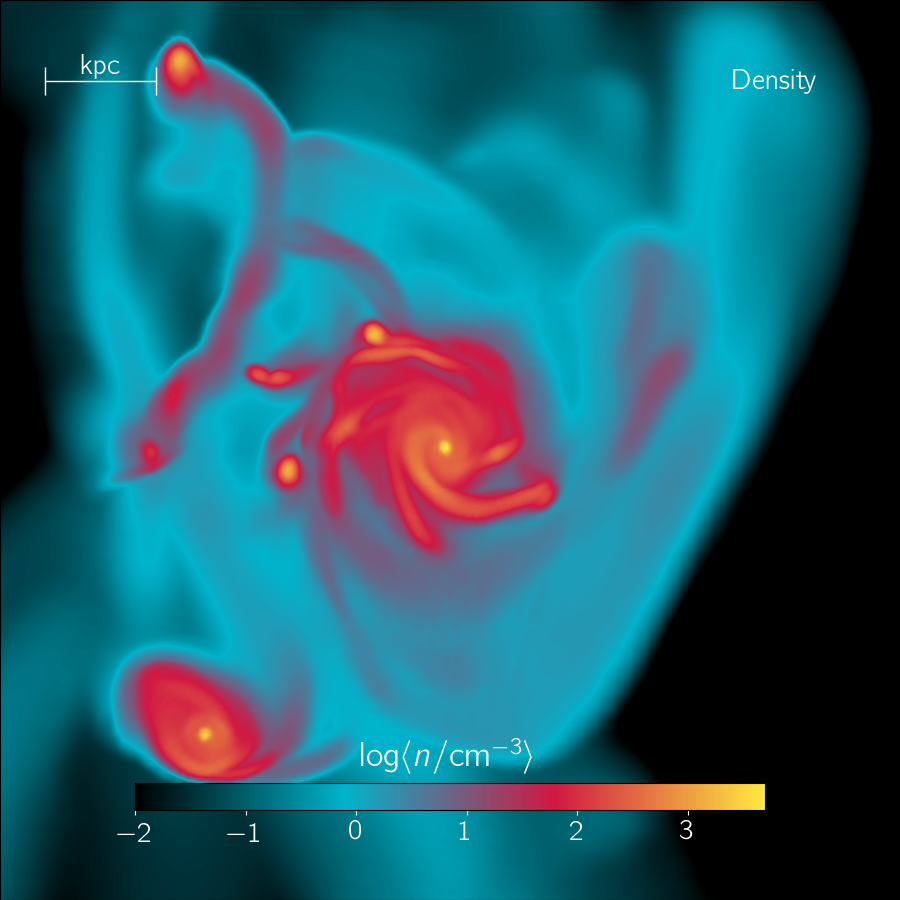}

\vspace{.3pt}
\includegraphics[width=0.32\textwidth]{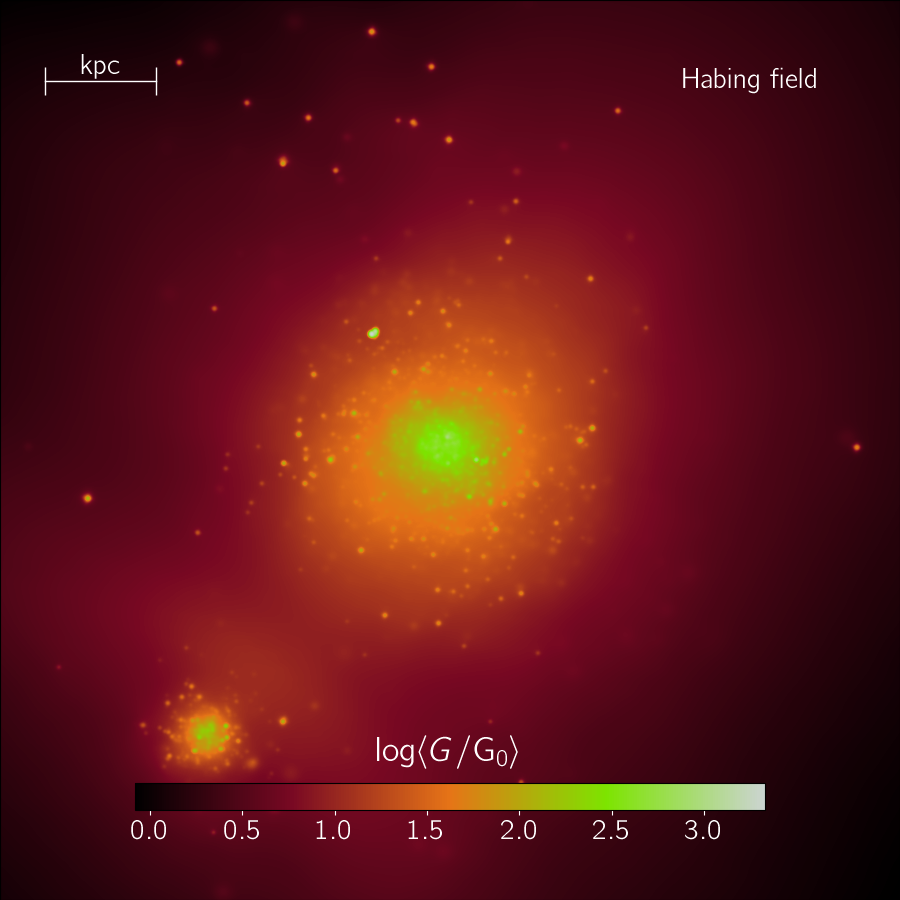}
\includegraphics[width=0.32\textwidth]{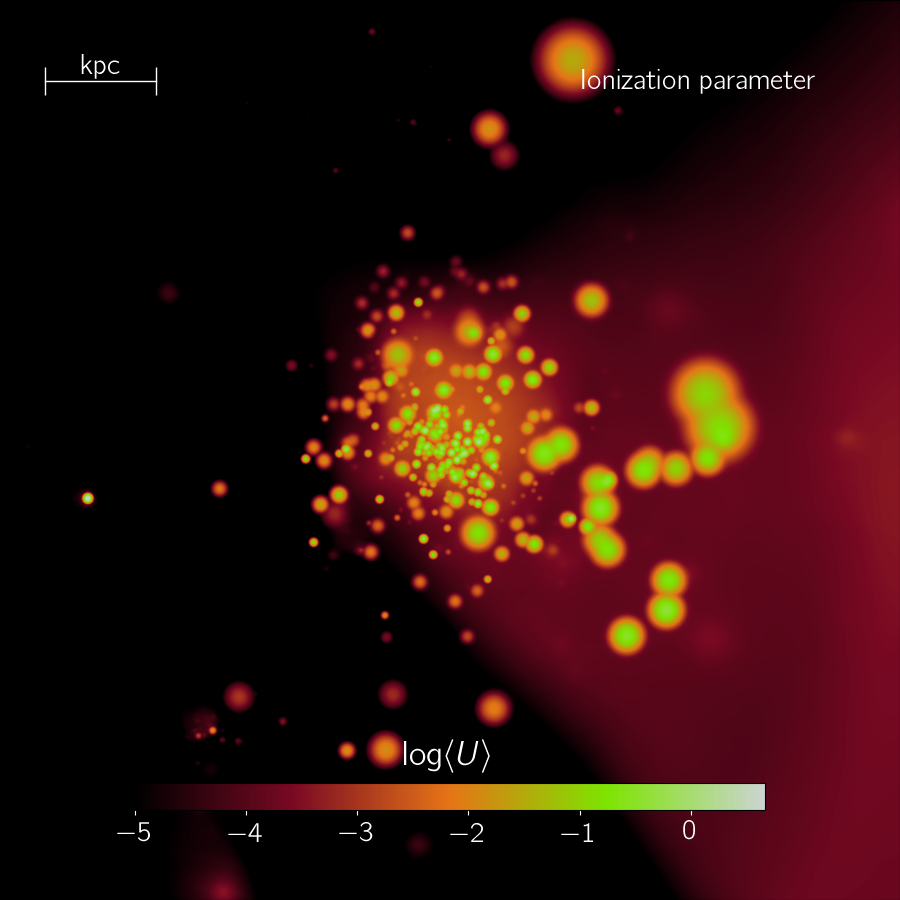}
\includegraphics[width=0.32\textwidth]{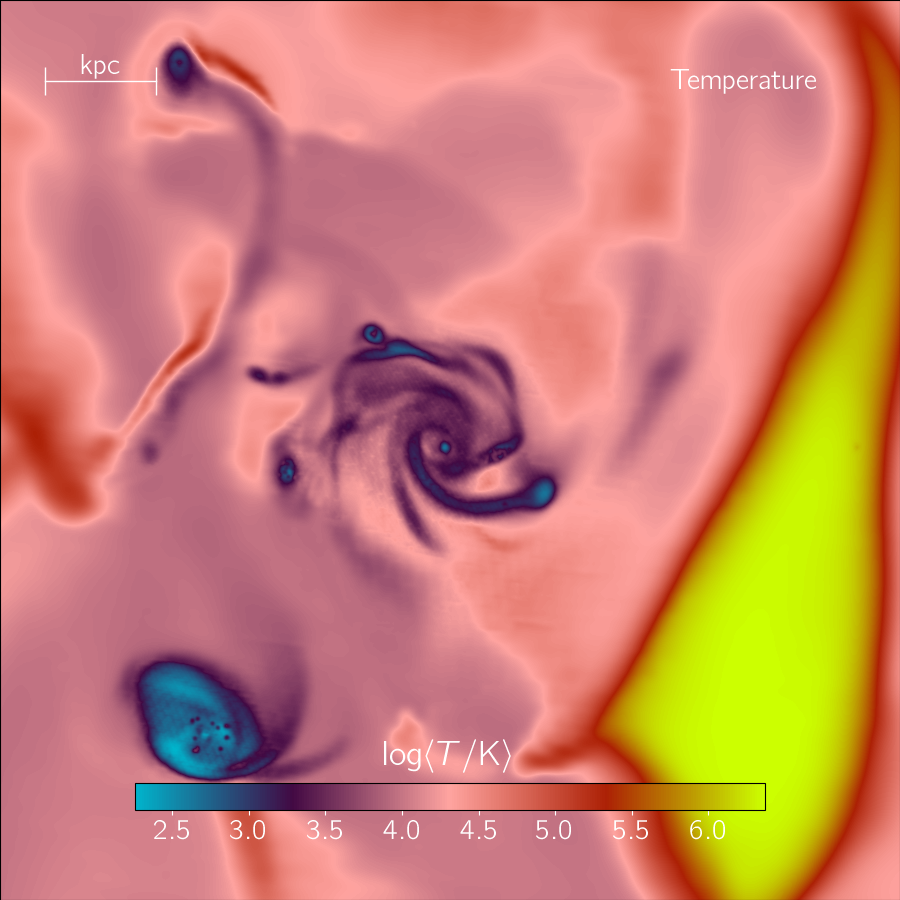}

\vspace{.3pt}
\includegraphics[width=0.32\textwidth]{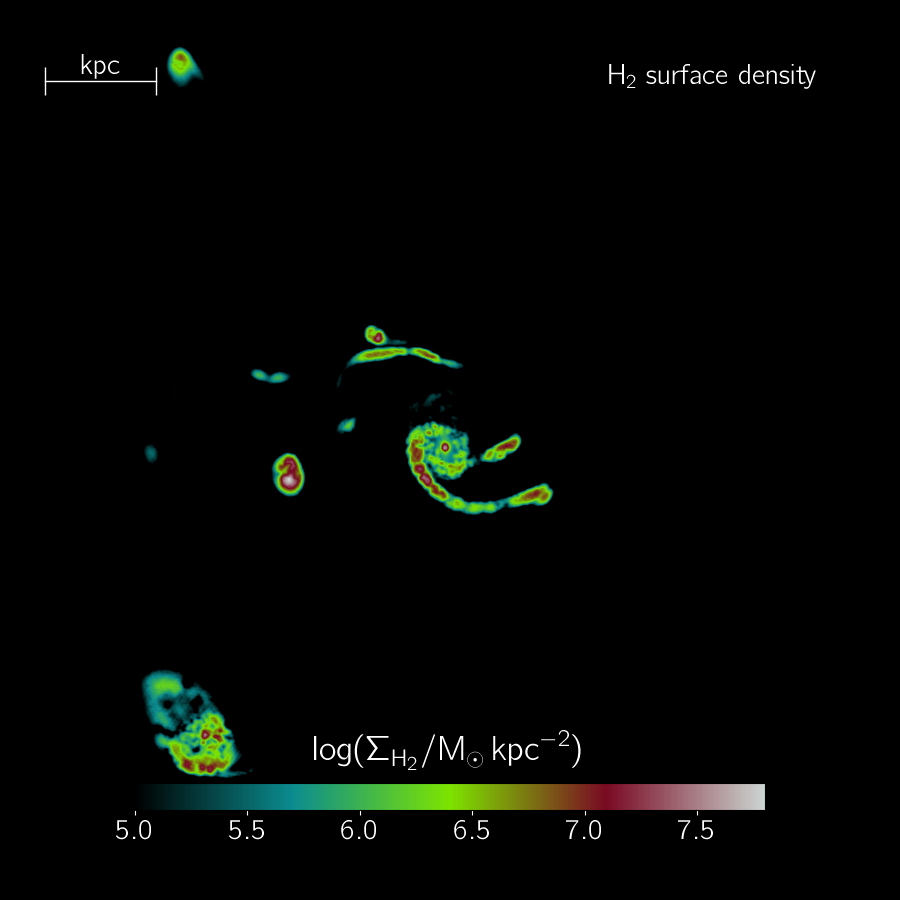}
\includegraphics[width=0.32\textwidth]{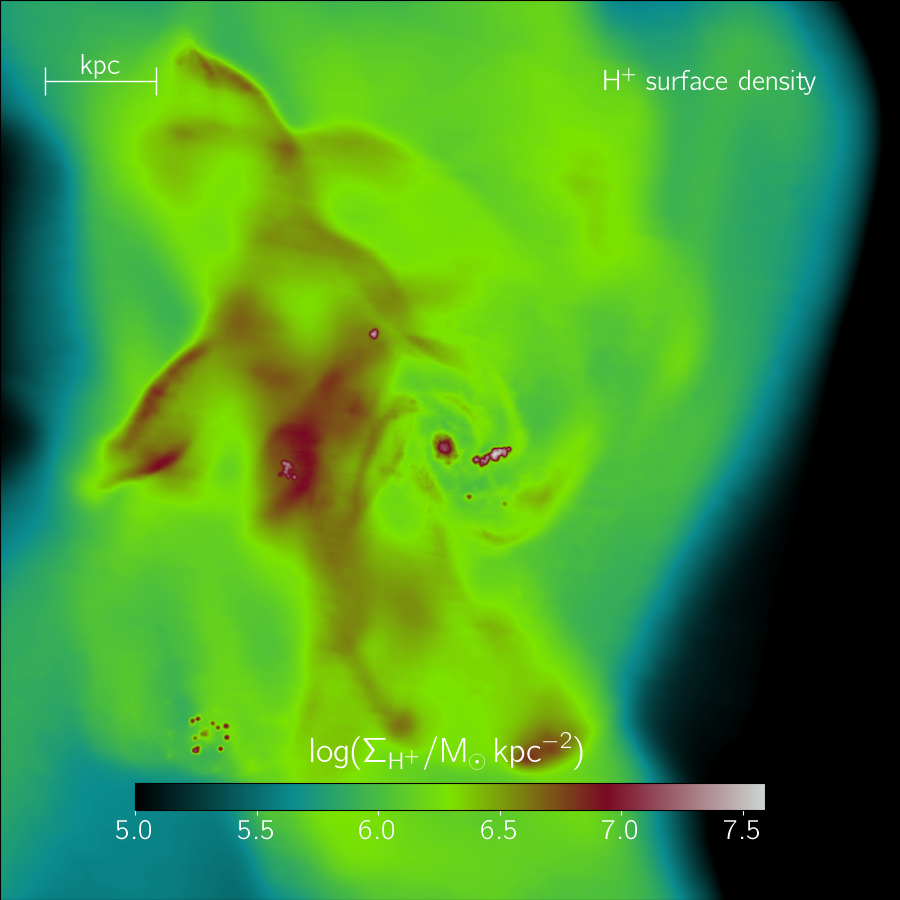}
\includegraphics[width=0.32\textwidth]{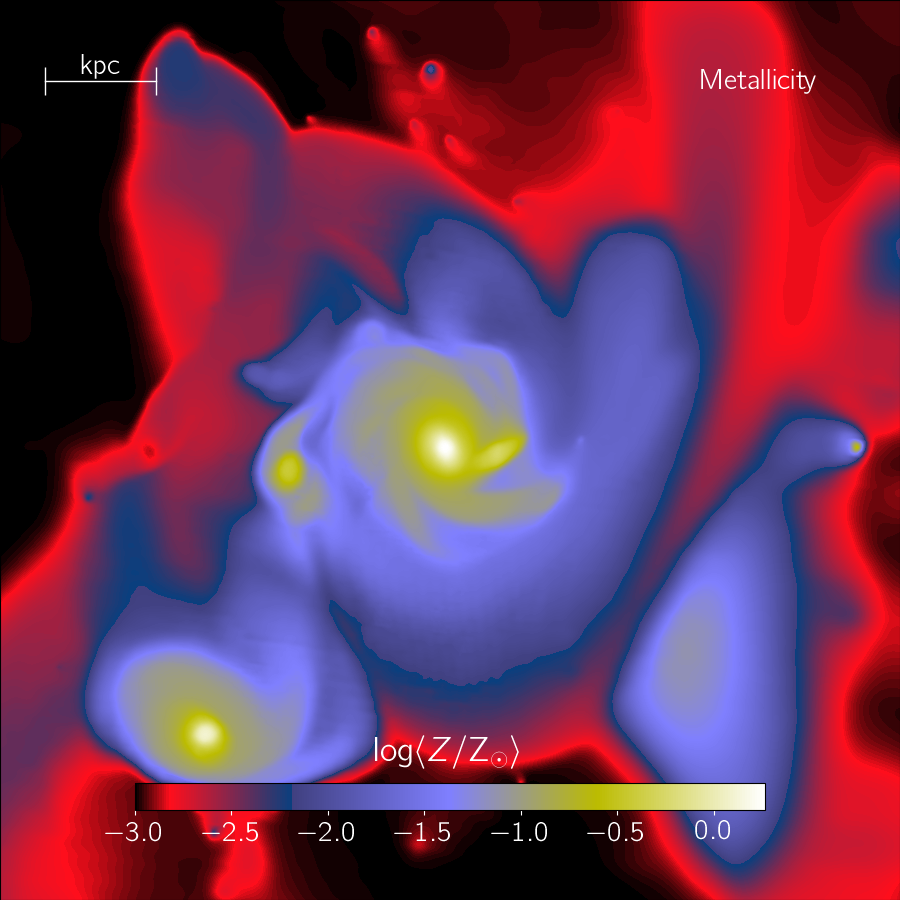}

\caption{
Portrait of Freesia at $z= 8$, when the galaxy has an age $t_\star \simeq 409\,\myr$, $M_{\star} \simeq 4.2\times 10^9 \msun$, and $\SFR \simeq (11.5\pm 1.8)\,\msunyr$.
The galaxy is seen face-on in a $\simeq (8.2 \,\kpc)^{2}$ field of view (FOV).
In the upper row we plot the stellar mass surface density ($\Sigma_{\star}$), star formation rate surface density ($\dot{\Sigma}_{\star}$), and the gas density ($n$).
In the middle row we show the Habing field ($G$), ionisation parameter ($U$) and the gas temperature.
In the bottom row molecular content ($\Sigma_{\rm H2}$), ionised hydrogen ($\Sigma_{\rm H^{+}}$), and gas metallicity ($Z$) are reported.
Note that $n$, $Z$ and $T$ are mass-weighted averages along the selected line of sight (l.o.s.); $G$ and $U$ are averaged by photon number; surface densities are integrated along the l.o.s.; $\dot{\Sigma}_{\star}$ accounts for star formed in the last $10\,\myr$.
\label{fig_overview_general}}
\end{figure*}

\begin{figure*}
\includegraphics[width=0.485\textwidth]{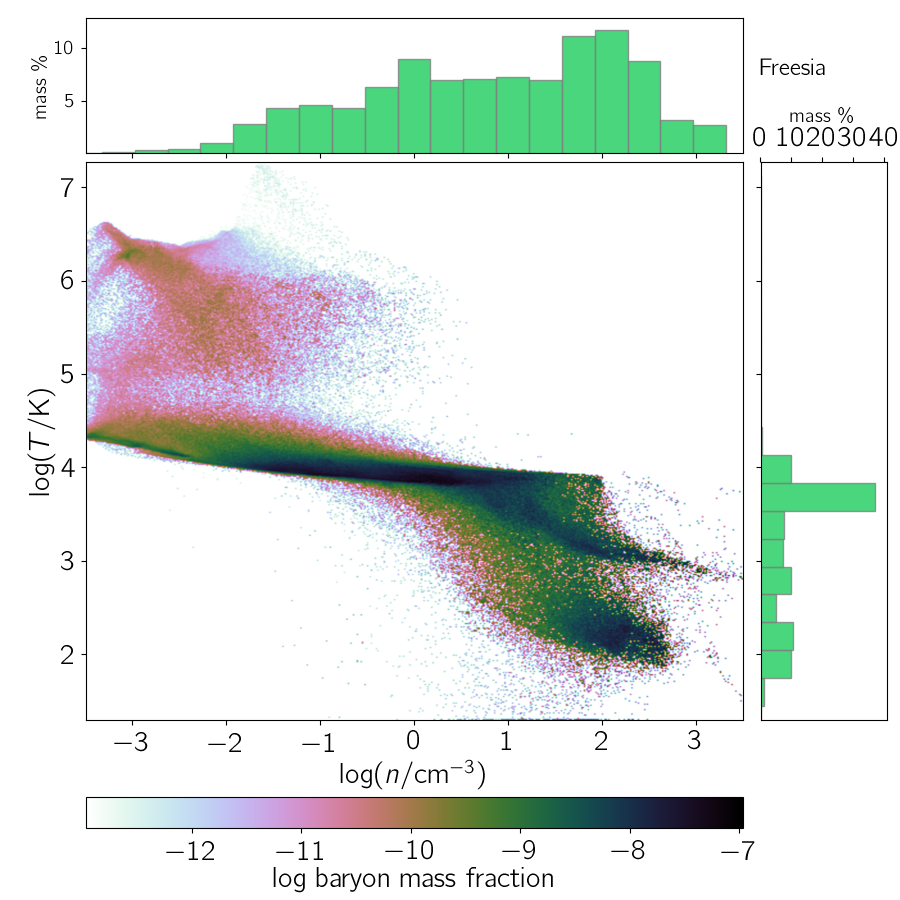}
\includegraphics[width=0.485\textwidth]{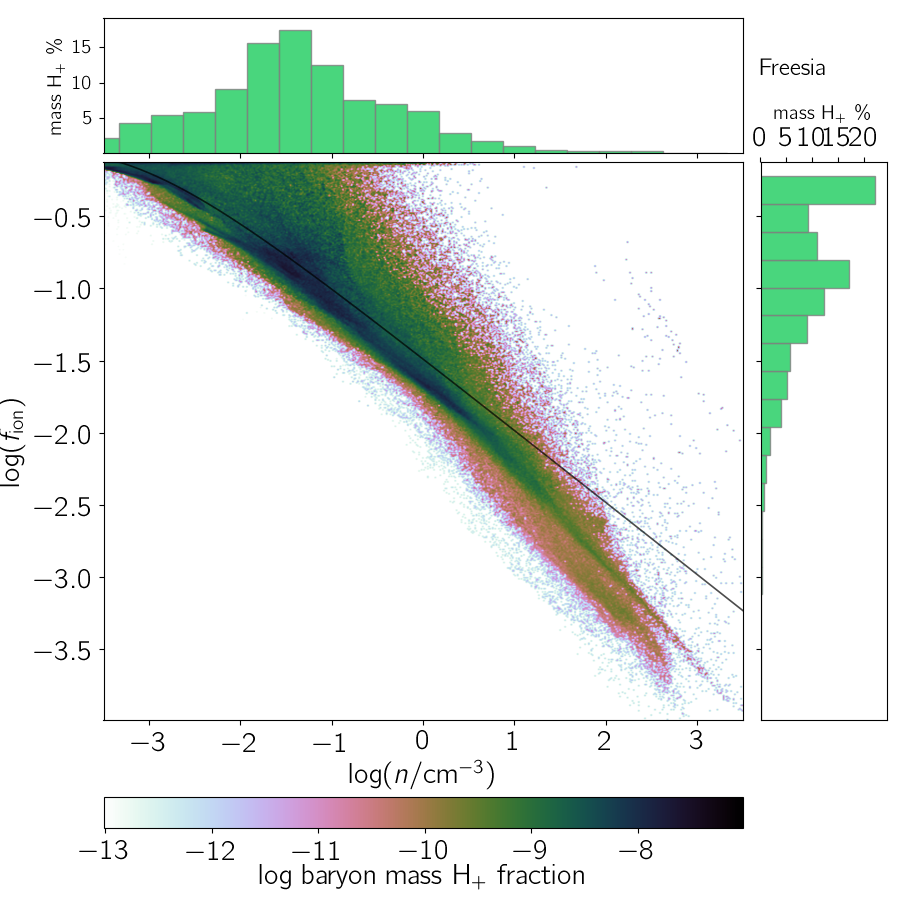}
\vspace{.3pt}

\includegraphics[width=0.485\textwidth]{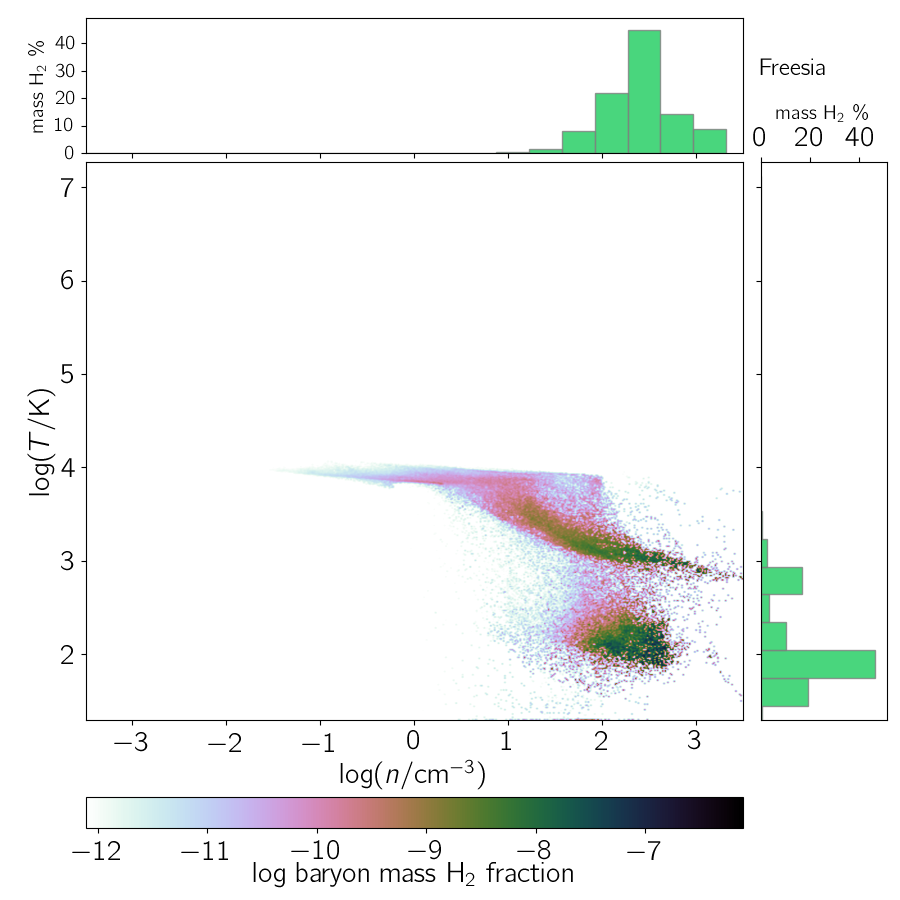}
\includegraphics[width=0.485\textwidth]{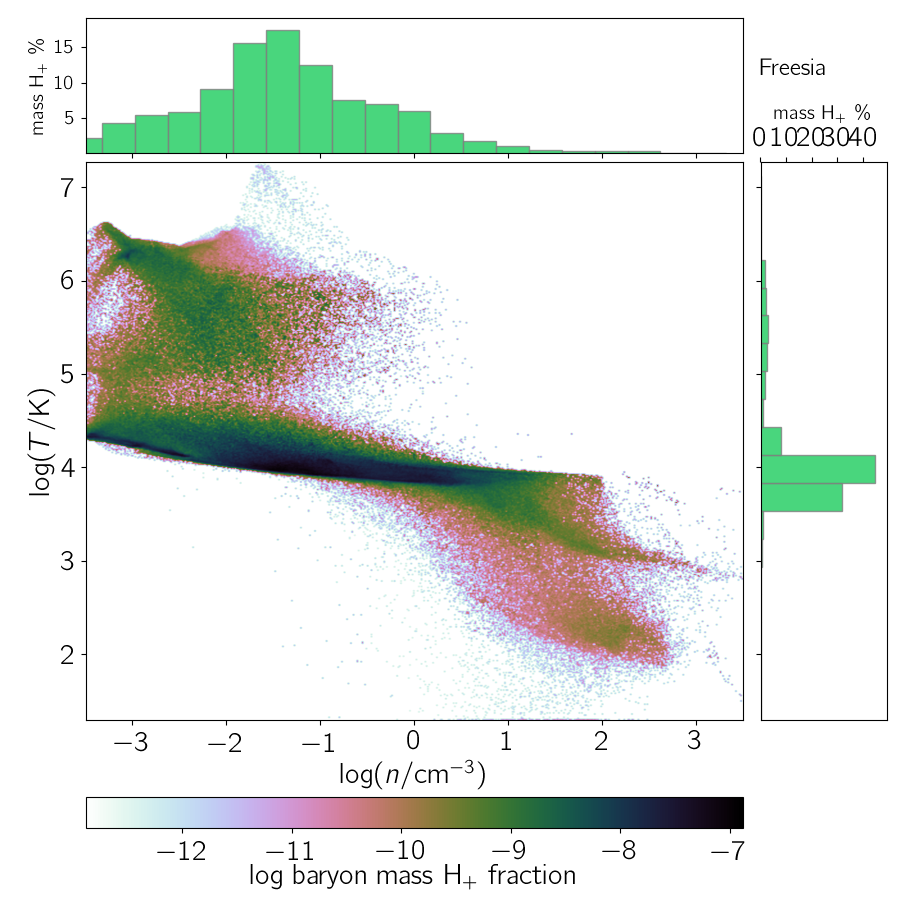}
\caption{
Phase-diagrams of the gas in Freesia. The phase-diagrams (or probability density function) in the density-temperature ($n$-$T$) plane are weighted by total gas mass ({\bf upper panel}), molecular hydrogen mass ($M_{\rm H2}$, {\bf lower left panel}) and ionised hydrogen mass ($M_{\rm H+}$ {\bf lower right panel}). The phase-diagram in the density-ionised fraction ($f_{\rm ion}$-$n$) plane is weighted by $M_{\rm H+}$ ({\bf upper right panel}).
These phase-diagrams account for the gas in a cubic region with side $8.2 \,\kpc$ centred on the galaxy, i.e. as the FOV shown in Fig. \ref{fig_overview_general}.
Each phase-diagram is plotted by normalising to unity the integral 2D integral. For each phase-diagram we additionally plot with a coarse binning the two projection in each axis, and the normalisation is chosen such that the sum of the value in the bins is $100\%$.
The black solid line in the $f_{\rm ion}$-$n$ diagram indicates the collisional ionisation for a gas at $T=10^4\rm K$
\label{fig_eos_chemistry}
}
\end{figure*}

\subsection{Galaxy formation histories}

The formation history of the sample of galaxies in our simulation is plotted in Fig. \ref{fig_time_evol_stars}, where we show the stellar mass build up ($M_{\star}$) and the star formation history (SFR, left panel) as a function of age ($t_\star$) and redshift ($z$, upper axis).

Freesia is the most massive galaxy in the sample. At $z=8$, it is hosted by a halo of mass $M_{\rm h}\simeq 10^{11}\msun$; its age is $t_\star \simeq 409\,\myr$, it has a stellar mass $M_{\star} \simeq 4.2\times 10^9 \msun$, and an instantaneous $\SFR \simeq (11.5\pm 1.8)\,\msunyr$, where the error is given by the variance in the last $10\,\myr$. The star formation shows variations on timescales of $\simeq 30\,\myr$, peaking up to ${\rm SFR}\simeq 30\,\msunyr$; overall the evolution is similar to our previous simulation without radiative transfer \citep[][]{pallottini:2017althaea}, with stellar mass differences of the order $5\%$, and variations in the SFR mostly due to the stochasticity of the star formation prescription (Sec. \ref{sec_model_stars}, see eq. \ref{eq_mean_poisson_sfr}).

Along with Freesia, there is a sample of eleven more galaxies in the simulated region. They have distance from Freesia that has mean (variance) of $56.9\, \kpc$ ($21.6\,\kpc$); as they are are at least two virial radii away ($r_{\rm vir}\sim 12\,\kpc$) at this redshift, we do not label them as satellites. Two of these galaxies are relatively massive, with $M_\star \simeq 5\times 10^8\msun$ and $\SFR\simeq 5\msunyr$, and they are younger ($t_\star\simeq 300\,\myr$) than Freesia. The other nine are smaller ($5\times 10^7\msun \lsim M_\star \lsim 10^8\msun$), with lower star formation rates ($\SFR\lsim \msunyr$) and they are typically much younger ($t_\star\lsim 150\,\myr$). {\commento Such small galaxies are hosted in $M_{\rm dm}\lsim 10^{9}$ dark matter haloes: their $M_{\rm dm}-M_{\star}$ relation has a large scatter, that is compatible with results from other theoretical works, which consider larger sample of galaxies \citep{xu:2016apj} and zoom-in simulations focusing on smaller galaxies evolved with higher mass resolution \citep{jeon:2015,jeon:2019}.}

As we had already noted in \citet[][]{pallottini:2017althaea}, most of these objects were not present in our previous simulations without radiative transfer, as their star formation was suppressed; this was entailed by our assumption on the spatially uniform ISRF. In the present work, the \HH~formation and hence star formation is not suppressed in objects that are located sufficiently far away from Freesia, as a consequence of flux dilution and attenuation.

In the remaining part of the Sec. along with Sec. \ref{sec_emission_prop} we focus on the properties of Freesia at $z=8$. In Sec. \ref{sec_obs_comparison}, the other systems are considered in the analysis.

\subsection{Freesia structural properties}\label{sec_freesia_structure}

A face-on representation of the key structural properties of Freesia is shown in Fig. \ref{fig_overview_general}. Freesia has two stellar components -- \quotes{A} and \quotes{B} -- separated by $\simeq 4\,\kpc$, with Freesia-A containing about $\simeq 85\%$ of the total stellar mass and dominating the star formation rate ($90\%$). Both components are highly concentrated, with effective radius of about $\sim 200\,\rm pc$ in both cases; they show stellar surface density peaks with $\Sigma_{\star}\simeq 5\times 10^{11}\surfd$, that are surrounded by a stellar halo with low surface density ($\Sigma_{\star}\simeq 10^{7}\surfd$), that is likely due to the tidal interaction of the components. Note that some of these stellar clusters have formed recently; while they give a negligible contribution to the total SFR, they can be important in the ionising photon budget.

Looking at the gas density, Freesia-A has a spiral structure with arms characterised by a gas density $10^{2}\lsim n/\cc\lsim 10^{3}$; Freesia-B reaches similar densities, but it has a more uniform disk structure, because of its lower mass prevent the development of arms. The only other dense ($n\simeq 10^2\cc$) structure is likely an unstable filament located $\simeq 2.5\,\kpc$ north-west of Freesia-A. These three components are embedded in a lower density medium ($n\simeq 5\,\cc$), with very low density ($\simeq 10^{-2}\cc$) shock-heated patches of gas\footnote{Note that $n\sim 10^{-2}\cc$ is a very low ISM density, but it is higher than the $\Delta = 10^{2}$ baryon overdensity that is usually selected to mark the edge of halos in cosmological simulations.}.

The average\footnote{Average values from the maps are typically quoted in the form ${\rm mean}\pm\, {\rm variance}$.} radiation field is $G = (7.9 \pm 23.1)\,G_{0}$ i.e. compatible with the assumption in \citet[][]{pallottini:2017althaea}, where $G = G_{0} ({\rm SFR}/\msunyr$) \citep[see also][]{behrens:2018}. Note that the variance of the radiation field is $3$ times the mean, as in the MW \citep{habing:1968,wolfire:2003apj}. In analogy with $\Sigma_\star$ and $\dot{\Sigma}_\star$ peaks, the Habing field has two maxima located at the centre of Freesia-A and Freesia-B and decreases radially from these locations because of flux dilution, as well as gas and dust absorption. While the spatial variance of the Habing field is small, various peaks ($\simeq 2\times 10^3 G_{0}$) are found in correspondence of stellar clusters, particularly in regions of recent star formation; pockets of gas with an enhanced local radiation can give an important contribution to line emission (Sec. \ref{sec_emission}, see Fig. \ref{fig_benchmark_cloudy}).

The ionisation field has an asymmetric structure and shows a larger variation, with a mean $U = (2\times 10^{-3} \pm 2\times 10^{-2})$; high values ($U\gsim 10^{-1}$) are co-located with recent star formation events, in particular in the region $2\,\kpc$ east of Freesia-A, where the gas density is low ($n\lsim 5\,\cc$). In the same region we can see that there is a trail of ionising photons that is leaving the galaxy with a conical shape.

The temperature map shows that the spiral arms of Freesia-A and the dense gas in Freesia-B are cold ($T\lsim 250 \,\rm K$) structures, blistered with warm ($T\sim 10^{4}\rm K$) spots, marking the presence of local radiative feedback. Shock heated regions ($T\sim 10^{5}\rm K$) that reach out from the two stellar components are caused by SN explosions, while the one west of Freesia-A is caused by accretion shocks.

The effect of local radiative feedback is more evident in the molecular and ionised hydrogen maps. The molecular material is concentrated in the spiral arms of Freesia-A and at the location of Freesia-B, with typical surface density peaks with $\Sigma_{\rm H2}\sim 10^{7.5}\surfd$ and sizes about $\simeq 100 - 400\, \rm pc$\footnote{See \citet{leung:2019} for a more complete analysis of individual MC properties found in our \code{serra} simulations.}. \Hp~regions also show similar values in the peaks of the distribution, i.e. $\sigma_{\rm H+}\sim 10^{7.5}\surfd$, but also enclose Freesia with a low surface density $\Sigma_{\rm H+}\sim 10^{6.5}\surfd$ halo component. In Freesia-B~the correspondence of \Hp~regions with spots of warm gas and local ionising field ($U\simeq 10^{-2}$) is particularly evident.

At this stage, Freesia is already mildly enriched, showing a metallicity of $Z\simeq 0.3\,\zsun$ in the dense region, with central peaks up to $Z\simeq 2\,\zsun$ in Freesia-A and Freesia-B. The surrounding gas is enriched at a mean $Z \simeq (0.02 \pm 0.06)\, \zsun$ up to $\sim 5\,\kpc$ from the stellar components.
{\commento Freesia has a deeper potential well ($M_{\rm dm}\simeq 10^{11}\msun$) with respect to the typical metal polluting galaxy ($M_{\rm dm}\lsim \times 10^{9}\msun$, c.f.r \citealt[][]{pallottini:2014sim}): indeed a $M_{\rm dm} \lsim 10^8 \msun$ halo is needed to have a galaxy with a mean $Z\lsim 0.05\, \zsun$ \citep[][see in particular Fig. 4 therein]{jeon:2015}}; thus SN shocks originating from Freesia are less effective in enriching the intergalactic medium, as only the galaxy immediate surroundings can be easily accessed.

\subsection{Thermo-chemical structure}

To analyse the thermo-chemical structure of the gas we look at the density-temperature phase-diagrams, i.e. mass-weighted probability density functions (PDF) in the $n$-$T$ plane. In Fig. \ref{fig_eos_chemistry} we plot the phase-diagrams weighted by the total gas mass, the molecular gas mass, and the ionised component for the material within $\simeq 4.1 \,\kpc$ from Freesia.

Considering the total mass (upper left panel of Fig. \ref{fig_eos_chemistry}), we see that $\simeq 60\%$ of the gas is photoheated ($T\simeq 10^4 \rm K$), due to photo-electric heating on dust grains illuminated by the radiation generated by local sources contributing to the Habing field. Lower temperatures can be reached by gas at density $n\gsim 10\,\cc$, accounting for $\simeq 35\%$ of the mass budget. The remaining $\lsim 5\%$ is shock heated ($T\simeq 10^6 \rm K$) by SN explosions and by accretion. The density has two small peaks around $n\simeq 1\cc$ and $n\simeq 10^{2}\cc$ that are superimposed to a roughly flat distribution. Overall, the total gas phase-diagram is similar to \citet{pallottini:2017althaea}; this is expected since in the former work a uniform ISRF is assumed, and in Freesia we find that the Habing field is rather uniform few $\kpc$ away from the stellar component.

The double-peaked nature of the density distribution is due to the presence of the molecular and ionised components, as it is clear from the lower panels of Fig. \ref{fig_eos_chemistry}.
\HH~is concentrated at high density (lower left panel of Fig. \ref{fig_eos_chemistry}), with the peak at $n\simeq 5\times 10^2 \cc$. The diagram features a prominent peak at $T\simeq 3\times 10^2\rm K$ and a less pronounced one around $T\simeq 10^3\rm K$. The presence of the latter is linked to the formation of \HH~in gas with $Z\sim 10^{-3}\zsun$, i.e. the dust channel is disfavoured with respect to gas phase formation, which is enhanced at $T\simeq 10^3\rm K$.
Note that there is a low ($\lsim 1\%$) amount of molecular hydrogen at even higher temperatures, albeit at lower ($n\lsim 10\,\cc$) densities. This gas is partially molecular and its presence is possible because of shielding from local LW sources. Recall that at this redshift \HH~lines fall into the spectral window of SPICA \citep{spinoglio:2017,egami:2018}, thus in the future it will be possible to test the presence of such relatively high temperature ($T\sim 10^3$) \HH, as the 0-0 S(1) at $17\,\mu\rm m$ and 0-0 S(0) at $28\,\mu\rm m$ are in the optimal sensitivity window of the instrument.

\Hp~is responsible for the low density peak that is seen in the total gas diagram (lower right panel of Fig. \ref{fig_eos_chemistry}). The bulk ($\simeq 70\%$) of ionised gas is centred at $n\simeq 5\times 10^{-2}\cc$ and it is photoionised at a temperature $T\simeq 10^4 \rm K$. Out of the total \Hp, $\simeq 25\%$ of the gas is in a shock heated state and mostly collisionally ionised. The remaining $\simeq 5\%$ of the ionised gas has typical densities $n\gsim 10\,\cc$ and is only partially ionised.

The ionisation structure is better appreciated by looking at upper right panel of Fig. \ref{fig_eos_chemistry}, that shows the phase-diagram of density vs ionised fraction, i.e. $f_{\rm ion}\equiv n_{\rm H+}/(n_{\rm H}+n_{\rm H+}+2\,n_{\rm H2})$. In Freesia, the ionised fraction decreases with density roughly as $f_{\rm ion}\propto n^{-0.5}$, with a dispersion of order of 0.3. The gas is found to be fully ionised ($f_{\rm ion}\simeq 1$) only in low density regions ($n\lsim 1 \cc$), while in potentially molecular regions ($n\sim 5\times 10^{2} \cc$) $f_{\rm ion} \simeq 10^{-3}$. Such is likely a spurious result deriving from unresolved high density \Hp~regions (see Sec.s \ref{sec_model_radiation} and \ref{sec_emission}).

Radiation gives a non-negligible contribution to the ionised fraction both in high and low density regions. Assuming local thermodynamical equilibrium, the ionisation fraction due to collisions can be written as $f_{\rm ion}^{\rm coll}=\sqrt{(1+\xi)^{2}-1} -\xi$, where $\xi = \Gamma_{\rm H}/(2 n \alpha_{\rm rec})$, with $\Gamma_{\rm H}=\Gamma_{\rm H}(T)$ and $\alpha_{\rm rec}=\alpha_{\rm rec}(T)$ being the collisional ionisation and recombination rate, respectively \citep[see e.g.][]{dayal:2008mnras,pallottini:2014cgmh}.
In Fig. \ref{fig_eos_chemistry} we plot $f_{\rm ion}^{\rm coll}$ for $T\lsim 10^{4}\rm K$ as a black solid line. $f_{\rm ion}$ is compatible with collisions only when $n\lsim 10^{-2}\cc$, while for progressively higher density the contributions from radiation becomes dominant, in particular considering that $f_{\rm ion}^{\rm coll} = 0$ for $T\lsim 10^{3}\rm K$, that is the typical temperature of the $n\gsim 10^2\cc$ gas.

\section{FIR emission properties}\label{sec_emission_prop}

\begin{figure*}
\includegraphics[width=0.32\textwidth]{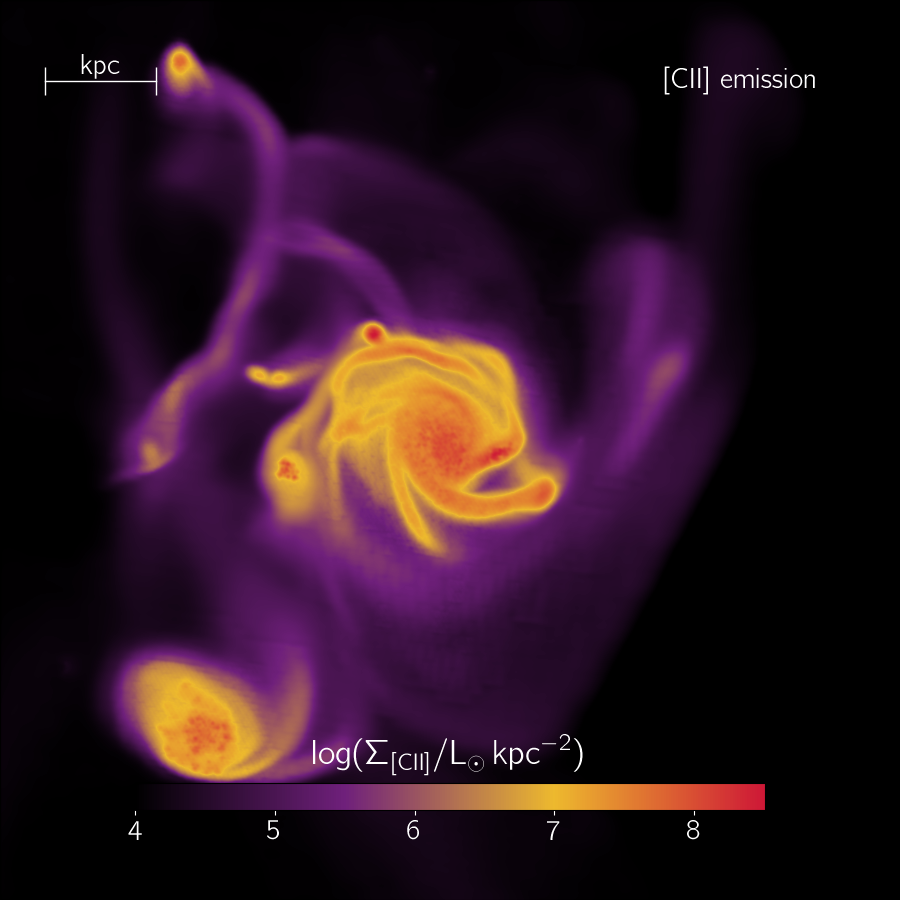}
\includegraphics[width=0.32\textwidth]{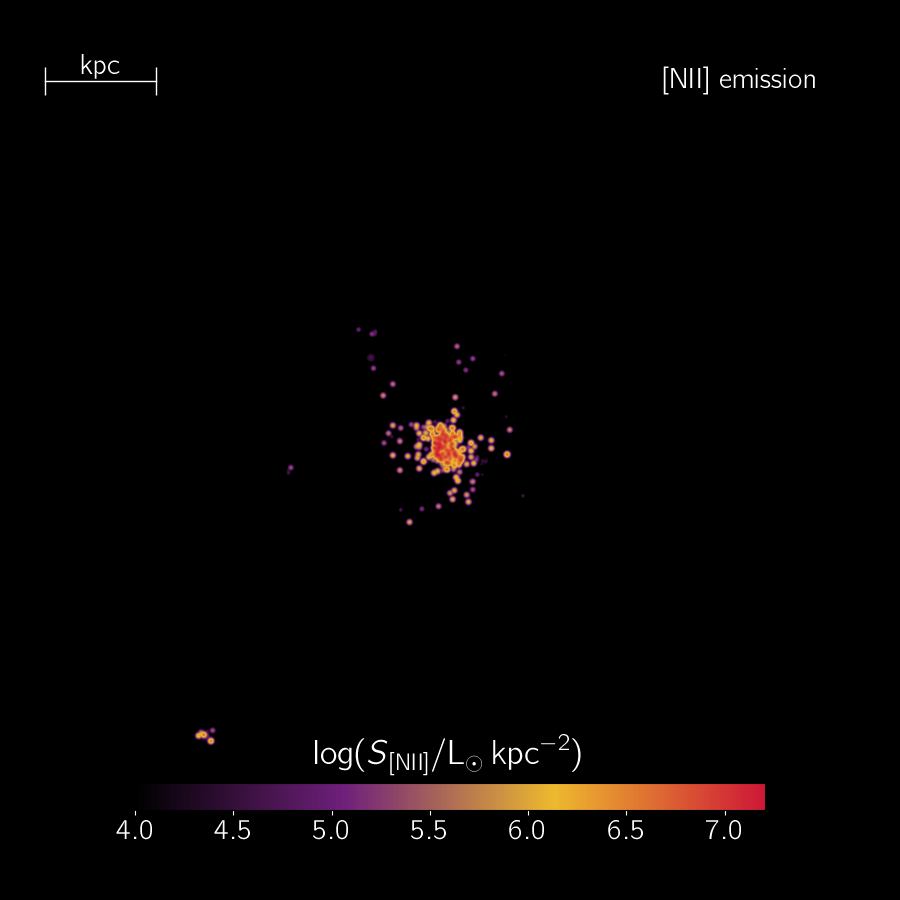}
\includegraphics[width=0.32\textwidth]{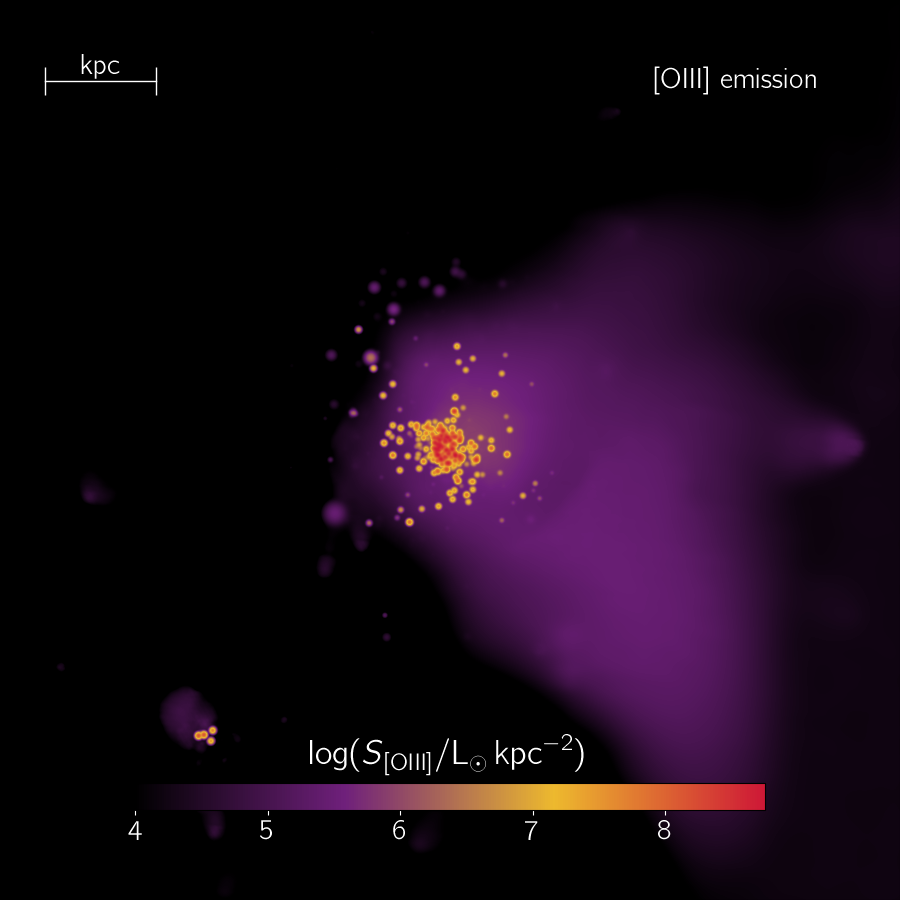}
\vspace{.3pt}

\includegraphics[width=0.32\textwidth]{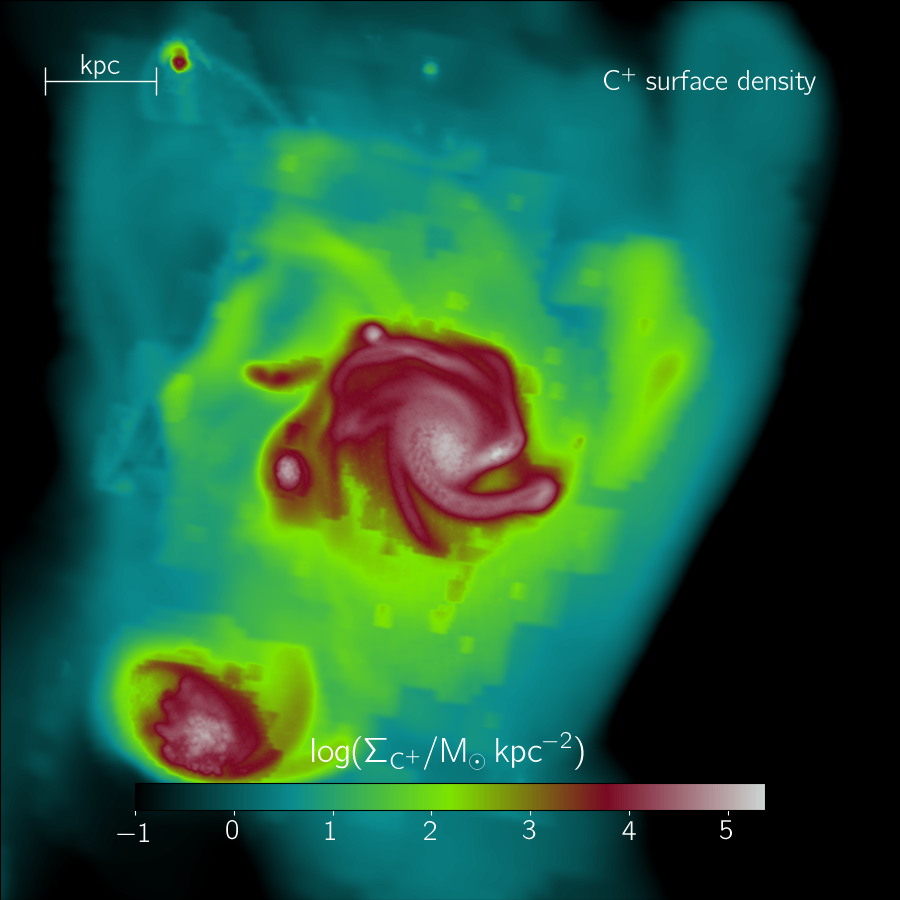}
\includegraphics[width=0.32\textwidth]{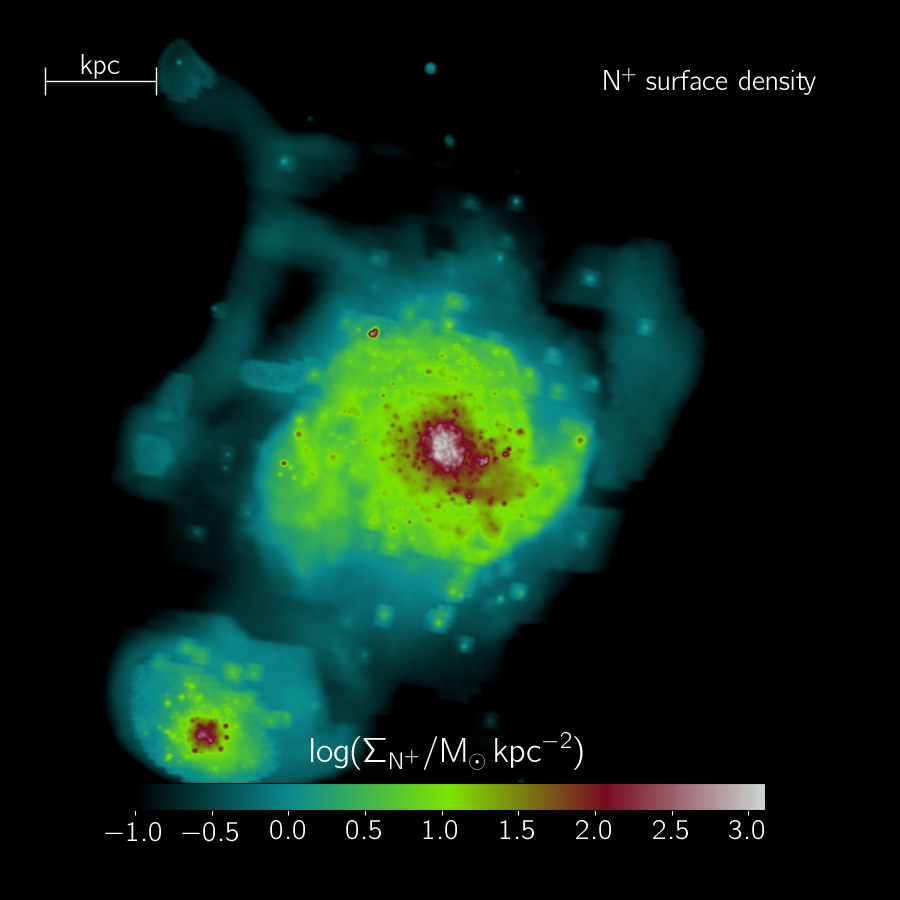}
\includegraphics[width=0.32\textwidth]{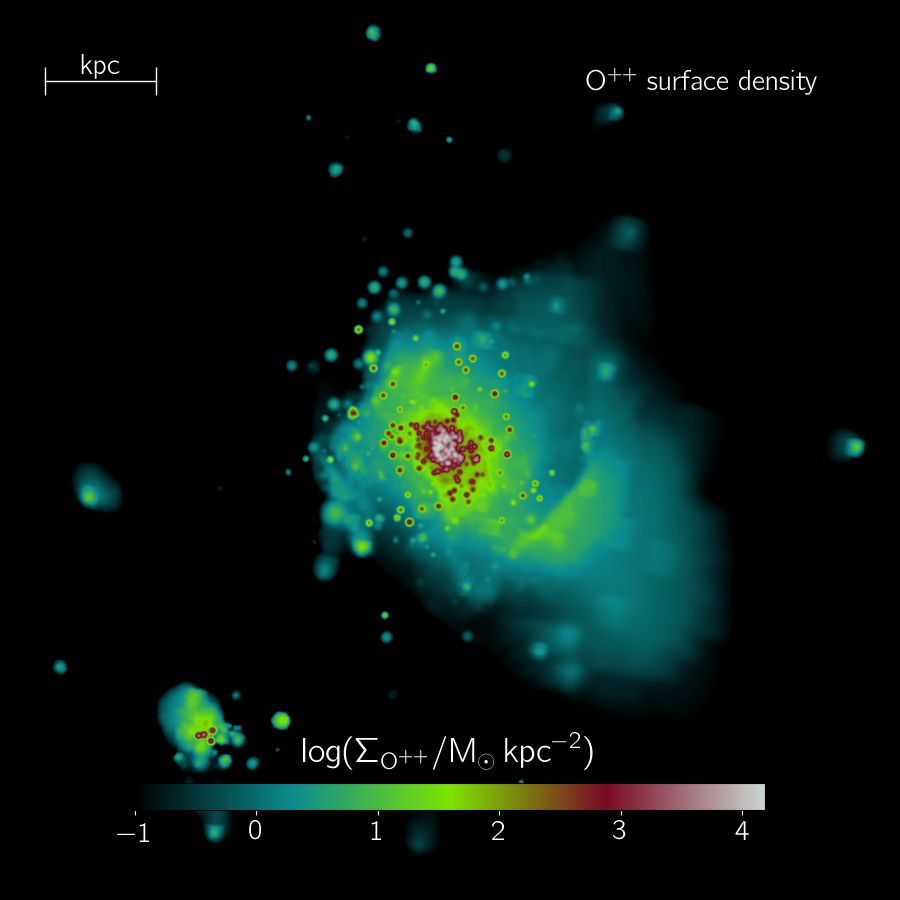}
\caption{
Far infrared (FIR) emission lines and corresponding ions in Freesia.
In the upper row we show the surface brightness for \CII~($\SCII$, {\bf left}), \NII~($\SNII$, {\bf centre}), \OIII~($\SOIII$, {\bf right}).
In the lower row we show the surface density for \CIIion~($\SigmaCII$, {\bf left}), \NIIion~($\SigmaNII$, {\bf centre}), \OIIIion~($\SigmaOIII$, {\bf right}).
The FOV is the same shown in Fig. \ref{fig_overview_general} and the integrated luminosities can be found in Tab. \ref{tab_summary_emission}.
\label{fig_map_emission}
}
\end{figure*}

\begin{table}
\centering
\begin{tabular}{llllll}
 \hline
 line  & $L$             & shift     & width ($\sigma_v$) & offset & radius\\
 ~     & $[\lsun]$       & [$\kms$]  & [$\kms$]           & [kpc]  & [kpc] \\ 
 \hline
 \CII  & $7.73 \times 10^7$ &  -             & 93.0     & -      & 1.54\\
 \NII  & $5.33 \times 10^5$ &  138.1         & 163.0    & 0.52   & 0.50\\
 \OIII & $2.07 \times 10^7$ &  121.1         & 163.3    & 0.65   & 0.85\\
\end{tabular}
\caption{Summary of the FIR emission line properties of Freesia. Emission line maps are given in Fig. \ref{fig_map_emission}, the spectra are plotted in Fig. \ref{fig_spectra}. Offset and radius are calculated from the location of the emission weighted mean and variance of the emission maps, respectively.
\label{tab_summary_emission}
}
\end{table}

\begin{figure*}
\includegraphics[width=0.485\textwidth]{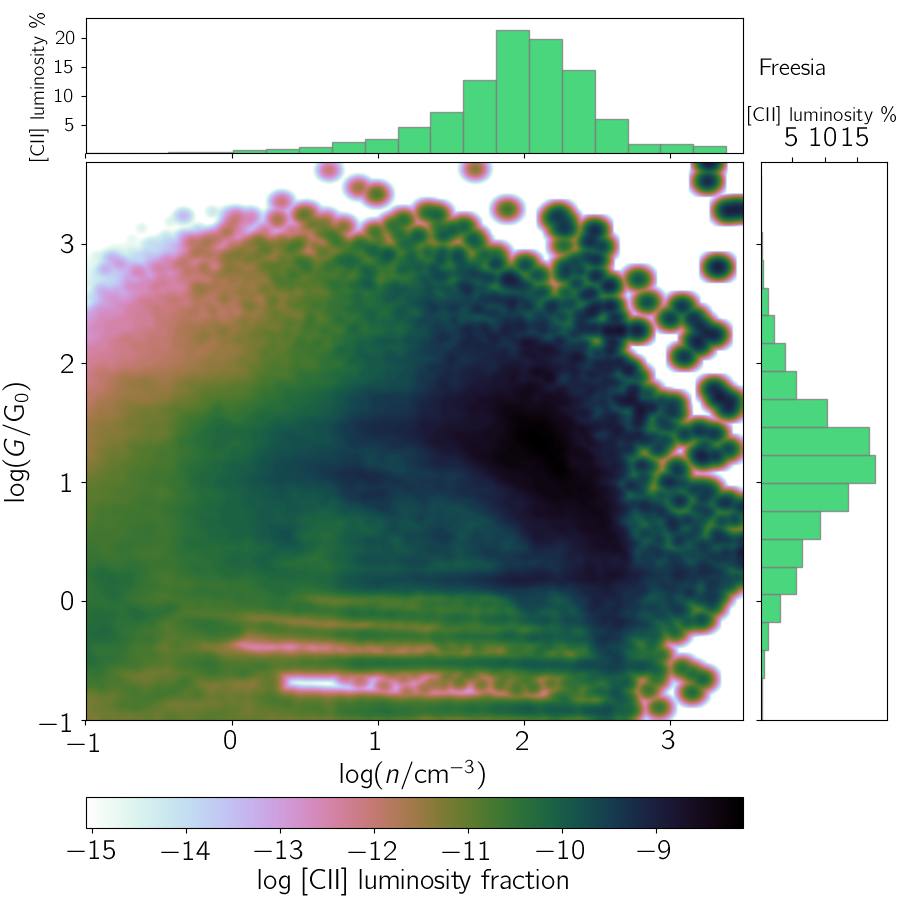}
\includegraphics[width=0.485\textwidth]{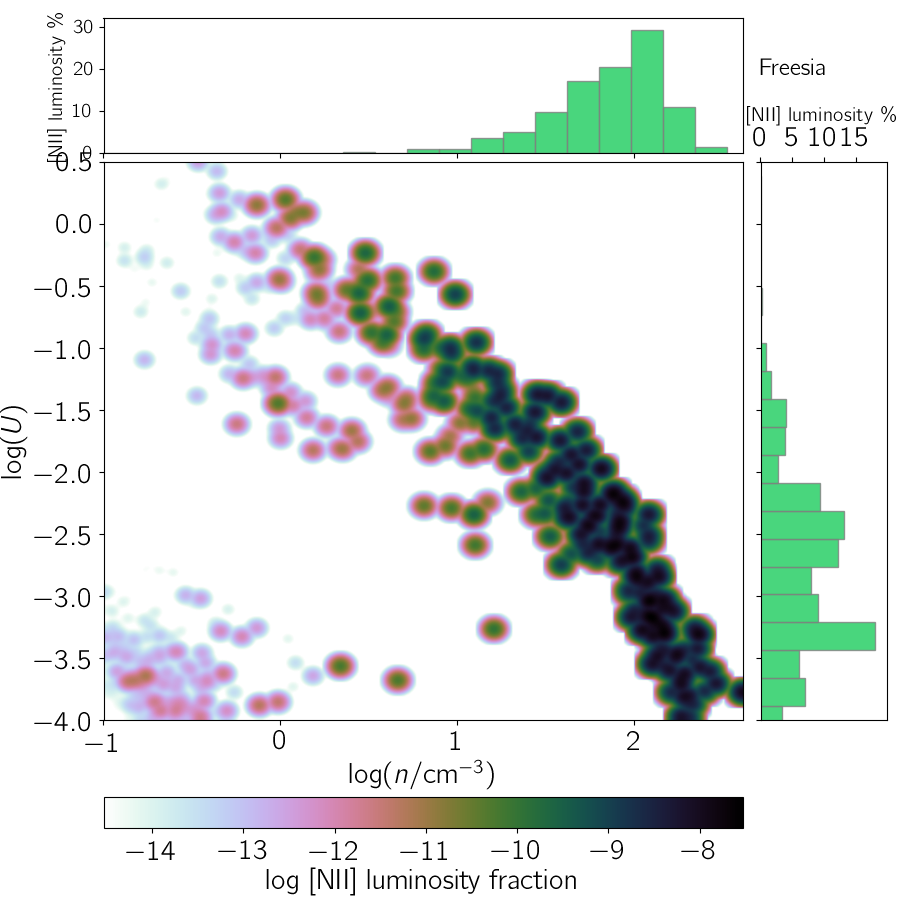}
\vspace{.3pt}

\includegraphics[width=0.485\textwidth]{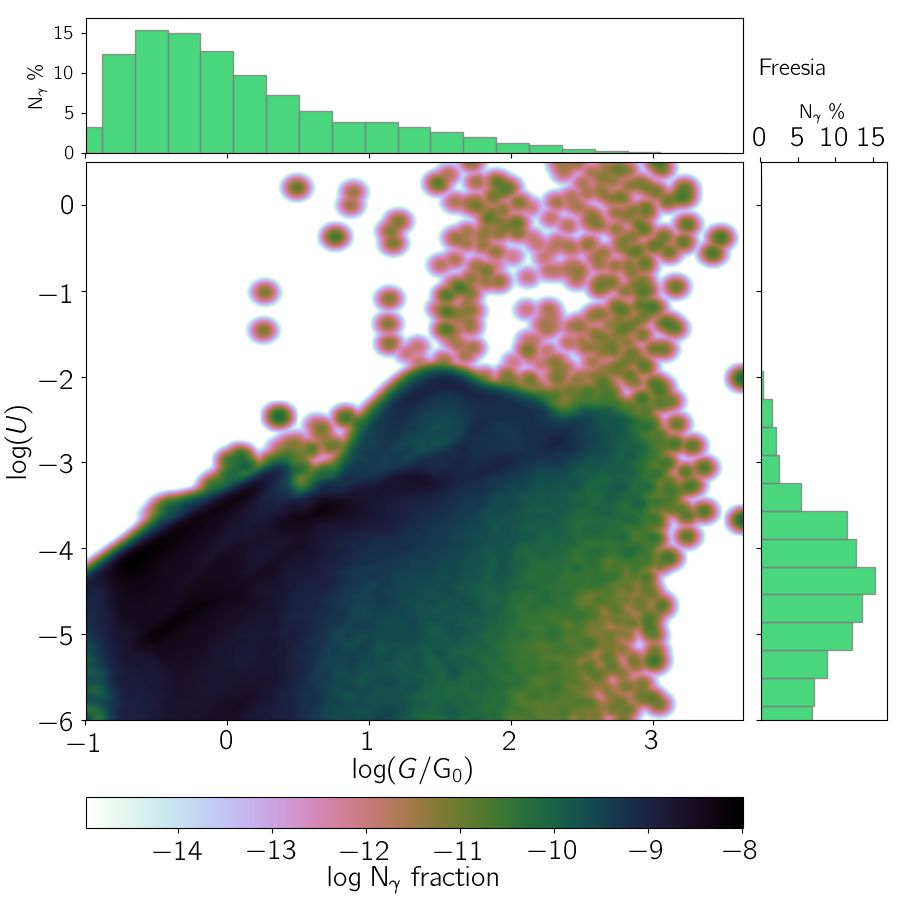}
\includegraphics[width=0.485\textwidth]{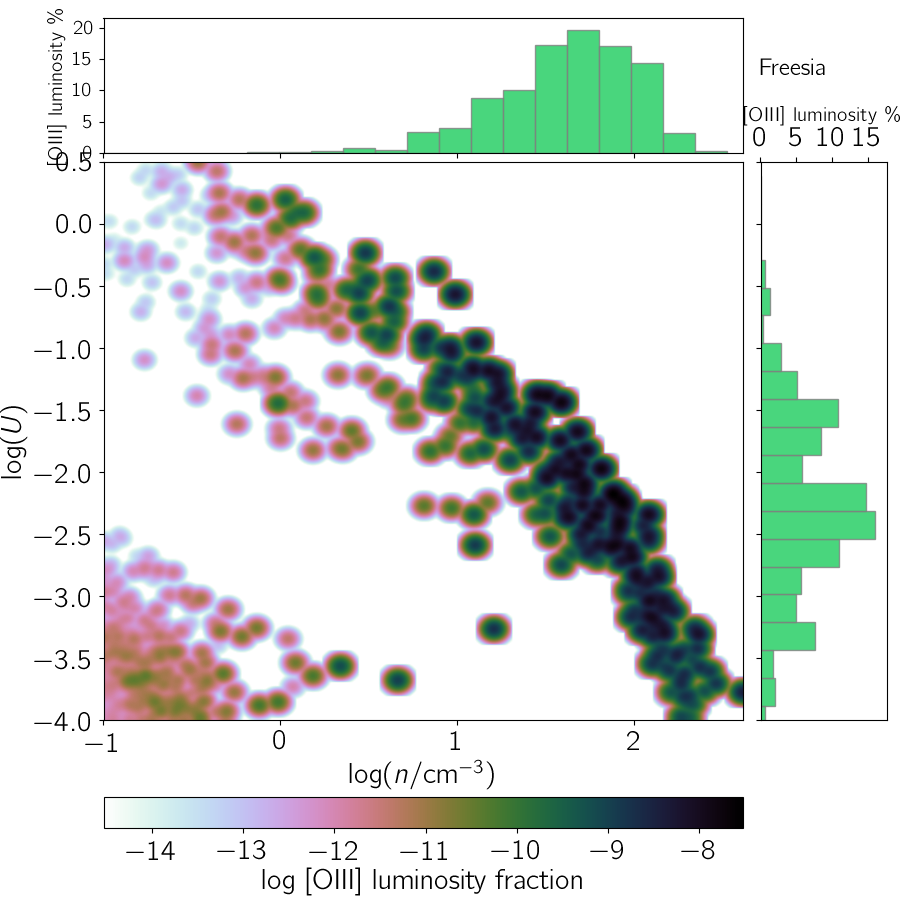}
\caption{
Phase-diagrams of the emission in Freesia. The luminosity weighted distribution for the \CII~in the $n$-$G$ plane ({\bf upper left}), \NII~in the $n$-$U$ plane ({\bf upper right}), and \OIII~in the $n$-$U$ plane ({\bf lower right}).
In the {\bf lower left} panel we plot the phase-diagram of $n$-$U$ weighted by total number of photon ($N_\gamma$).
Notation and references are is for the gas phase-diagrams (in Fig. \ref{fig_eos_chemistry}), however -- for visualisation sake -- a Gaussian smoothing has been applied to the calculation of the 2D probability.
\label{fig_eos_emission}
}
\end{figure*}

\begin{figure}
\includegraphics[width=0.485\textwidth]{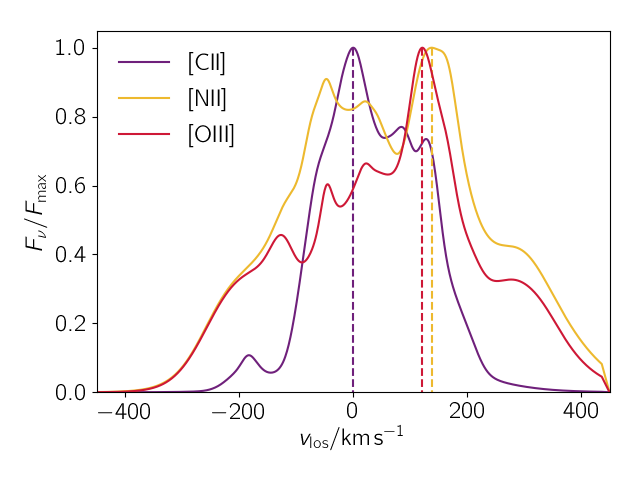}
\caption{
Freesia \CII, \NII, and \OIII~line spectra. For each emission line, the spectrum ($F_{\nu}$) is normalised to the peak flux ($F_{\rm max}$) and plotted as a function of the l.o.s. velocity ($v_{los}$). Spectra are extracted from the FOV given in Fig. \ref{fig_map_emission} and the $v_{los}$ is calculated according to the map orientation. $v_{los}$ is centred on the \CII~peak, and vertical dashed lines highlight the velocity peak of each spectrum.
A summary of the properties can be found in Tab. \ref{tab_summary_emission}.
\label{fig_spectra}
}
\end{figure}

In this work we study the following FIR emission lines: \CII~$158\mu$ ${}^2 P_{3/2}\rightarrow {}^2 P_{1/2}$, \NII~$122\mu$ ${}^3 P_{2}\rightarrow {}^3 P_{1}$, and \OIII~$88\mu$ ${}^3 P_{1}\rightarrow {}^3 P_{0}$, and we analyse the abundance and spatial distribution of their relative ions, \CIIion~(ionisation potential $11.26\, \rm eV$),~\OIIIion ($35.11\, \rm eV$), and \NIIion~($14.53\, \rm eV$).

\subsection{Imaging of the FIR emission and ions}

In Fig. \ref{fig_map_emission} we show the emission maps of Freesia~in \CII, \NII, and \OIII. As a reference, the properties are summarised in Tab. \ref{tab_summary_emission}.

The brightest line is \CII, with $L_{\rm CII} \simeq 7.7\times 10^{7}\lsun$. The two main peaks ($\SCII\simeq 5\times 10^7\surfl$) are spread over the Freesia-A~and Freesia-B, with extents of the order of $1.5\,\kpc$. The only other prominent structure is the high density filament ($n\simeq 10^2\cc$) North-West of Freesia, featuring a lower brightness -- $\SCII\simeq 5\times 10^6\surfl$ -- due to the lower metal content ($Z\simeq 5\times 10^{-2}\zsun$, cfr. with Fig. \ref{fig_overview_general}). The bright spots are embedded in a faint -- $\SCII\simeq 5\times 10^4\surfl$ -- halo that marks the extent of region that has been metal enriched by Freesia. Note that such diffuse halo in Freesia gives only a small contribution to the \CII: the $1.5\,\kpc$ extension of the emitting region extension is mainly determined by the presence of the two stellar components.

The emission from ions with higher ionisation state has a different morphology than \CII. Both \NII~and \OIII~show are less extended than \CII. For \NII~and \OIII, the luminosity of Freesia-A~is a factor $\gsim 10$ larger than Freesia-B, while for \CII~ the factor is $\simeq 4$. Since they trace similar material (see later Fig. \ref{fig_eos_emission}), the two lines have cospatial emission peaks, located in high density ($n\gsim 10^2\cc$) ionised regions that blister the disk of Freesia and enclose a total size $\lsim 0.5 \kpc$.
In particular, the \NII~line has a low luminosity, $L_{\rm NII} \simeq 5.3\times 10^5\lsun$. This is due to the smaller cooling efficiency with respect to the other lines \citep[][]{dalgarno:1972}: the maximum surface brightness is $\SNII\simeq 2\times 10^7\surfl$, roughly one order of magnitude smaller than \CII~and \OIII.
While the \OIII~surface brightness is higher than \CII~($\SOIII\simeq 5\times 10^8\surfl$), the smaller emitting region makes its total luminosity lower, i.e. $L_{\rm OIII} \simeq 2.1\times 10^7\lsun$. The \OIII~shows a more diffuse halo with surface brightness $\SOIII\simeq 5 \times 10^4\surfl$. However with respect to the \CII~halo the morphology is very different, since the \OIII~halo is confined in the east direction, in correspondence of the low ionisation field ($U\sim 10^{-4}$).

Along with emission lines, the corresponding ion surface densities ($\SigmaCII$, $\SigmaNII$, and $\SigmaOIII$) are plotted in Fig. \ref{fig_map_emission}. \CII~structure follows the \CIIion~ion morphology. \CIIion~is present in both diffuse and molecular material, and -- without ionising radiation -- we expect $\SigmaCII\propto n Z$ \citep[][]{ferrara:2019}. Since the Habing field is almost constant on these scales, a rough proportionality between the luminosity and the ion abundance is expected. In both Freesia-A~and Freesia-B the gas features a flat $\SigmaCII\simeq 10^5 \surfd$, that rapidly decreases to $\SigmaCII\lsim 10^2 \surfd$ in the diffuse halo. As the unstable filament north-west of the galaxy has a lower metal enrichment than the two star forming components, its \CIIion~abundance is similar to that of the halo.
\NIIion~and \OIIIion are similarly distributed and trace the ionised high density regions, as for the corresponding lines. In both cases, halos of low surface density material are present ($\Sigma\lsim 10 \surfd$), however the ISRF is not high enough in order for the corresponding lines to be emitted efficiently. Note that the order of magnitude difference in the surface density is mostly due to the difference between the mass abundance of the elements.

\subsection{FIR lines as a tracer of the ISM state}

Using the phase-diagrams we can analyze the emission structure of the ISM in Freesia. In Fig. \ref{fig_eos_emission} we plot the phase-diagrams weighted by the luminosity of \CII~($n$-$G$ plane), \NII~($n$-$U$ plane), and \OIII~($n$-$U$ plane).

The \CII~diagram shows that most of the contribution to its emission comes from gas with $n\simeq 160\, \cc$ illuminated by $G\simeq 20 \,G_{0}$, i.e. dense gas in the two star forming components embedded in the average radiation field; this peak spans roughly an order of magnitude in both axis.
Contribution from lower density gas ($n\lsim 10 \cc$) is present but subdominant ($\lsim 10\%$). Note that for a weak field ($G\lsim G_{0}$) at low densities ($n\lsim 10 \cc$), the contribution to the emission is also suppressed by the CMB \citep{dacunha:2013apj,vallini:2015,pallottini:2015cmb}. Overall these results are consistent with our previous findings \citep[][]{pallottini:2017dahlia,pallottini:2017althaea}, i.e. most of the \CII~emission is associated with material close to the molecular regions.
A peak at high density is expected from our benchmark (Sec. \ref{sec_emission}, in particular see Fig. \ref{fig_benchmark_cloudy}), which also shows that \CII~emission is favoured at relatively higher values of the Habing field ($G\gsim 5\times 10^2 G_{0}$). These regions are relatively rare in Freesia, as they are associated with star forming regions, thus in our galaxy the peak contribution comes from regions with a milder radiation field.

Similar distributions are found for \NII~and \OIII. Two types of emitting regions are present: i) a stripe with roughly $U\propto n^{-2}$, which accounts for most of the luminosity of both lines, and ii) a $\lsim 5\%$ contribution from diffuse ionised gas with $n\lsim 1\,\cc$ and $U\simeq 5\times 10^{-4}$.

In both cases the emission peaks in dense \Hp~regions. However, some differences are present: the \NII~peak is concentrated at $U \simeq (2\pm 0.1)\times 10^{-3}$ for gas with densities $n \simeq (95 \pm 40)\, \cc$, while \OIII~shows a larger range for the ionisation parameter, $U \simeq (6\pm 2)\times 10^{-3}$, and arises from gas with lower densities, $n \simeq (53 \pm 40)\, \cc$. Thus, additionally to the higher typical cooling efficiency, the higher \OIII~luminosity with respect to \NII~is also partially due to contribution from gas with $n\simeq 10\, cc$, which is more abundant then the $n\simeq 95\,\cc$ gas that dominates the \NII~line emission.

To analyse the spectral hardness of the radiation field, in Fig. \ref{fig_eos_emission} we also plot phase-diagram in the $G$-$U$ plane weighted by photon number ($N_\gamma$). Overall, the bulk of the ISRF is relatively soft, with $\langle U\rangle = 7\times 10^{-4} \pm 0.03$ and $\langle G\rangle = (15.0 \pm 71) G_0$, and has a trend roughly given by $U \propto (G/G_{0})^{1/2}$ for $G\lsim 10^{2} G_{0}$

The region in the phase-diagram corresponding to high $U$ and $G$ is characterised by young stellar clusters that have removed the gas from their surrounding, through their radiative and mechanical feedback (see also Fig. \ref{fig_overview_general}). This is motivated as follows: before absorption, a typical stellar SED would yield $G/G_0\sim U (n/\cc) 10^3$ (see Fig. \ref{fig_sed}); as $G\sim 10^3 G_0$ at most, the density in $U\sim 1$ regions must be $n\lsim 1\,\cc$, i.e. lower than the original $n\simeq 3\times 10^2\cc$ allowing the formation of stars.

Summarising, most of the radiation surrounding Freesia is non ionising, i.e. the escape fraction calculated on a sphere of radius of $4.1 \kpc$ is of order $\simeq 2\%$, in agreement with estimate from observations of lower redshift galaxies with similar brightness \citep[e.g.][]{bowens:2016,grazian:2017} and averaged values of simulated galaxy with similar masses \citep{xu:2016apj}. However, large variations are expected with different evolutionary stage and radii considered \citep{trebitsch:2017}; a detailed analysis is left for future works.

\subsection{Spatial offsets}

We have seen that while low (\CIIion) and high (\OIIIion) ionisation species have a roughly similar spatial distribution, their corresponding emission structure is very different as a result of the modulation imposed by the IRSF, which is roughly spatially uniform in the Habing band, and very patchy in ionising radiation.
This causes a spatial offsets between \CII~and \OIII~emitting regions. In Freesia there are two different kinds of spatial offsets. In Freesia-A~\CII~is extended and peaks at the edge of the disk, while \OIII~is concentrated at the center. This configuration results in a $\simeq 250 \rm pc$ offset for the \CII~and \OIII~lines arising from Freesia-A. In Freesia-B~ \CII~has luminosity $\sim 10^7 \lsun$, but $L_{\rm OIII}\lsim 10^5 \lsun$. While Freesia-B~would not be detected in \OIII~even with an extremely deep ALMA observation, its \CII~luminosity would move the center of the emission away from Freesia-A; thus a marginally resolved observation of the system would reveal an offset of $\simeq 2 \rm kpc$ between \CII~and~\OIII. This is similar to what is observed in BDF-3299 \citep[][in particular see Fig. 4.]{carniani:2017bdf3299}

We recall that in BDF-3299  $L_{\rm CII}/L_{\rm OIII} = 0.27 $ \citep[see Tab. 4 of][]{carniani:2017bdf3299}, while in Freesia we find $L_{\rm CII}/L_{\rm OIII} = 3.73$.
In Freesia, the \OIII~emission predominantly comes from $Z\simeq 0.5\, \zsun$ gas in Freesia-A; there, $\SCII/\SOIII \simeq 0.1$.
%
The \OIII~emission is limited to regions with an hard radiation field ($U\gsim10^{-2}$), while \CII~can excited by the mean UV ISRF ($G\simeq 7.9\,G_0$) in the diffuse medium surrounding Freesia-A and thus is emitted efficiently also from the material surrounding Freesia-A, yielding a ratio of total luminosity $L_{\rm CII}/L_{\rm OIII}<1$.
This difference between Freesia and BDF-3299 can be explained if the latter has a larger ionised region; however it might be due to a different configuration of the system: further investigation is needed to have achieve a full classification.

Interestingly, neither in Freesia nor in the other simulated galaxies we find situations in which $L_{\rm CII}/L_{\rm OIII}\ll 1$, as shown in observations where \OIII~is present but \CII~is undetected \citep{inoue:2016sci,laporte:2017apj}.
We note that the three simulated galaxies at $9 \leq z < 12$ presented in \citet[][see Fig. 11 therein]{katz:2019arXiv} are also \CII-dominated with typical values $L_{\rm CII}/L_{\rm OIII}\sim 10-50$.
At present, simulations are apparently unable to reproduce the observed low $L_{\rm CII}/L_{\rm OIII}$ ratios. This issue requires further study and we leave it for future, more specific analysis.

\subsection{Spectral shifts}

Spectral shifts between different FIR lines are present in Freesia. To quantify this effect, we build spectra ($F_{\nu}$) as a function of the line of sight (l.o.s.) velocity ($v_{los}$). Each gas patch along the line of sight gives a contribution corresponding to its luminosity, which is kernel weighted by a Gaussian centred at the peculiar velocity of the gas and with a width that accounts for the thermal and turbulent broadening. Full detail of the model are given in \citet{kohandel:2018}, along with an in-depth analysis of the kinematics of the \CII.

In Fig. \ref{fig_spectra} we plot the \CII, \NII, and \OIII~spectra (see Tab. \ref{tab_summary_emission} for reference).
The \CII~appears to be more peaked in velocity space, with a line width (second moment of the spectrum) $\sigma_v = 93.0\, \kms$. The \CII~ emission comes from both stellar components, with Freesia-A~providing a broader turbulent disk component with $|v_{\rm los}|\lsim 150\,\kms$, while Freesia-B~provides a concentrated contribution at $v_{\rm los} = 0$, that gives rise to the peak in the total spectrum. This is a rather common situation when multiple, possibly merging, components are present in the same system \citep[][]{kohandel:2018}. Note that only $\lsim 10\%$ of the emission comes from high velocity ($|v_{\rm los}|\gsim 150\,\kms$) gas, as most of the emission is concentrated in the dense ISM of the galaxy; only a small fraction possibly due to outflowing material, because of its lower metallicity \citep{gallerani:2016outflow}.
The \OIII~line is shifted with respect to \CII~by $v_{\rm los} = 138.14\, \kms$, roughly one third of what is observed for BDF3299 \citep[][]{carniani:2017bdf3299} as its emission is dominated by Freesia-A. In Freesia \OIII~has a $\sigma_v$ that is $1.75$ larger than \CII, and it additionally features more prominent high velocity wings at $|v_{\rm los}|\gsim 150 \,\kms$, that are due to the diffuse low surface brightness \OIII~halo in the West region (see Fig. \ref{fig_map_emission}). This possibly makes \OIII~ a better tracer of outflowing gas.
The spectrum of \NII~feature double peaks, due to both the contribution from Freesia-A~and Freesia-B, that are almost coincident with the location of \CII~and \OIII~peaks in the velocity space. A luminosity ratio \CII/\NII $\sim 1/100$ is expected also from theoretical works \citep[][]{vallini:2013mnras}. This makes \NII~observations very challenging.

Note that the spatial offsets and spectral shifts typically observed in high-$z$ galaxies are between UV/Ly-$\alpha$ and \CII; only in a handful of cases \OIII~is also available \citep[][see Fig. 6.]{carniani:2017bdf3299}. Such analysis requires additional post-processing work to compute the UV continuum and Ly-$\alpha$ radiative transfer, as in e.g. \citet{behrens:2019}. This is also deferred to a forthcoming paper \citep[][]{kohandel:2018}.

\section{On the [CII]-SFR relation}\label{sec_obs_comparison}

\begin{figure}
\includegraphics[width=0.485\textwidth]{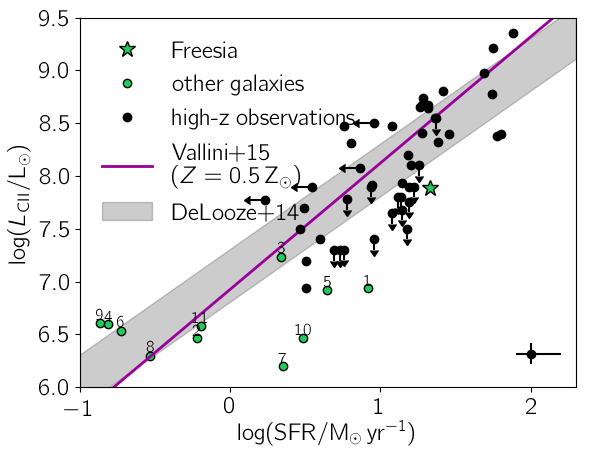}
\caption{Integrated SFR-\CII~relation. Freesia is marked with a star and simulated galaxies are indicated with green circles. Simulated galaxies are numbered in order to allow a better comparison with Fig. \ref{fig_sk_global}.
Along with the simulated point we plot ALMA observations of high-$z$ galaxies as re-analysed in \citet[][in particular see Tab. 2 and Fig. 4 therein]{carniani:2018}, with original data taken from \citet{barisic:2017,capak:2015arxiv,carniani:2017bdf3299,carniani:2018himiko,matthee:2017,jones+17,inoue:2016sci,ota:2014apj,schaerer:2015,smit:2018,willott:2015arxiv15} and \citet{kanekar:2013,bradac:2017,schaerer:2015,knudsen:2016arxiv} for lensed galaxies; contribution from the obscured SFR is included for objects where dust continuum measurements are available. The median error bar for the sample is plotted in the lower right corner.
With a gray band we plot the \citet{delooze:2014aa} relation for local galaxies.
Model from \citep{vallini:2015} for $Z = 0.5\,\zsun$ is plotted with a purple line \citep[for the analytical form see  eq. 1 of][]{pallottini:2015cmb}.
\label{fig_cii_sfr}
}
\end{figure}

\begin{figure}
\centering
\includegraphics[width=0.485\textwidth]{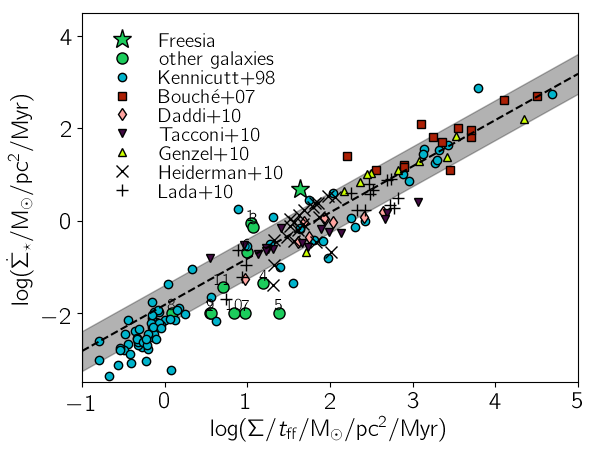}
\caption{
Comparison of the observed and simulated Kennicutt-Schmidt (KS) relation, i.e. star formation rate surface density ($\dot{\Sigma}_{\star}$) plotted against the gas density divided by free fall time ($\Sigma/t_{\rm ff}$).
Observation encompass data of single MCs \citep{heiderman:2010,lada:2010apj}, local unresolved galaxies \citep{kennicutt:1998apj}, and moderate redshift unresolved galaxies \citep{bouche2007apj,daddi:2010apj,daddi:2010b,tacconi:2010Natur,genzel:2010mnras}.
The correlation (dispersion) for the observation found by \citet[][see the text for details]{krumholz:2012apj} is plotted with a black dashed line (grey shaded region).
Simulation data for Freesia and the other galaxies is unresolved, i.e. it is obtained by degrading the $\dot{\Sigma}_{\star}$ and $\Sigma/t_{\rm ff}$ images with a gaussian filter with width equal to the size of the galaxy.
\label{fig_sk_global}
}
\end{figure}

\begin{figure*}
\includegraphics[width=0.32\textwidth]{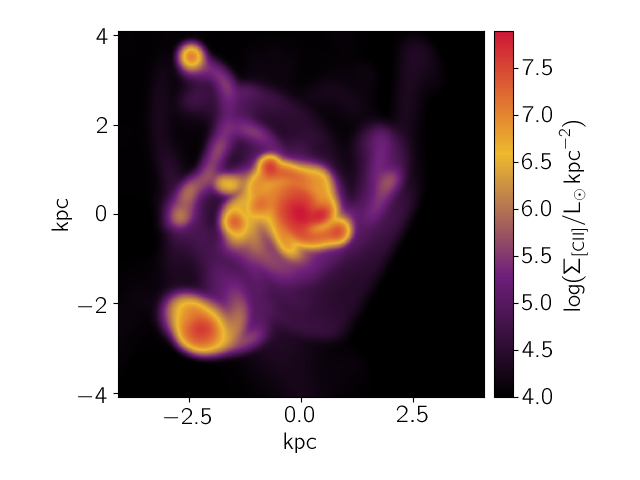}
\includegraphics[width=0.32\textwidth]{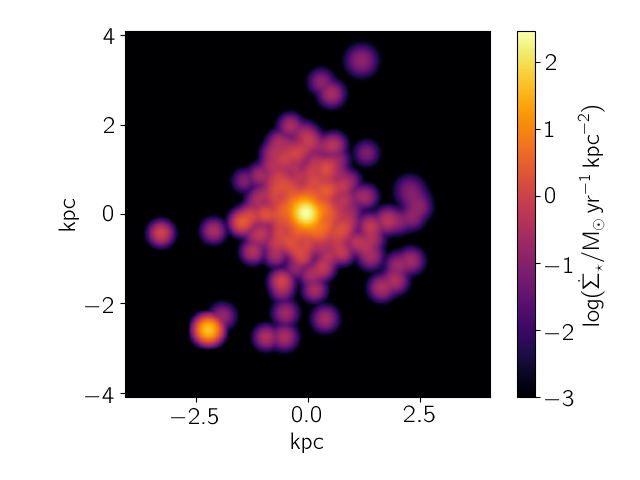}
\includegraphics[width=0.32\textwidth]{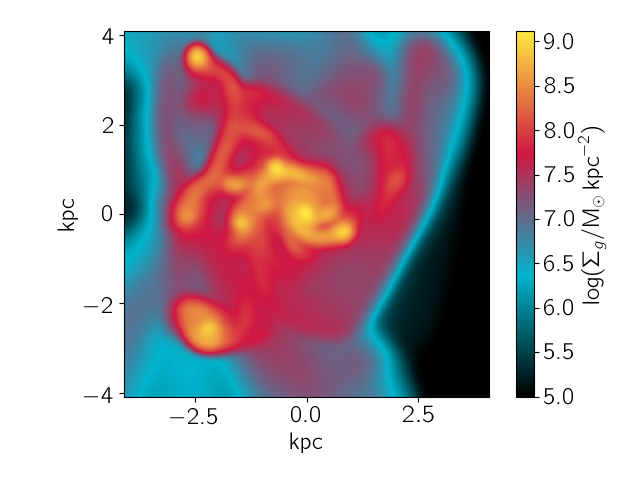}
\vspace{.3pt}

\includegraphics[width=0.32\textwidth]{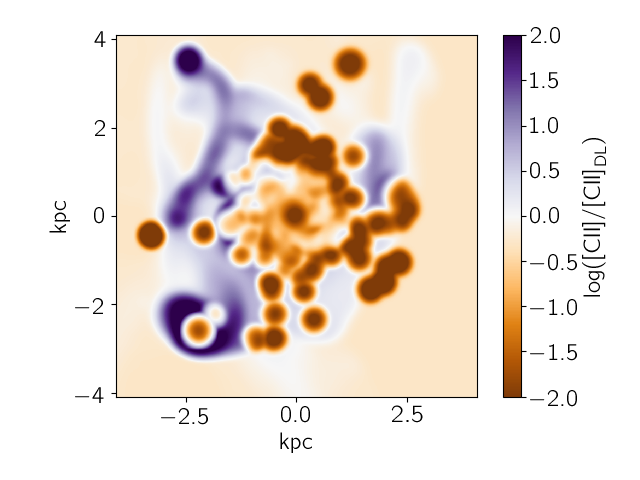}
\includegraphics[width=0.32\textwidth]{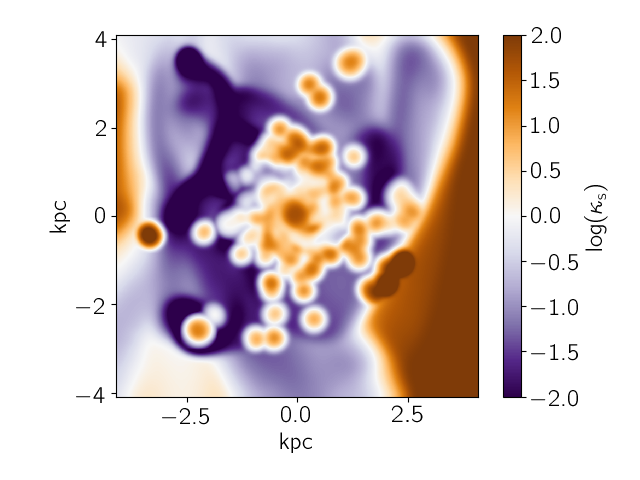}
\includegraphics[width=0.32\textwidth]{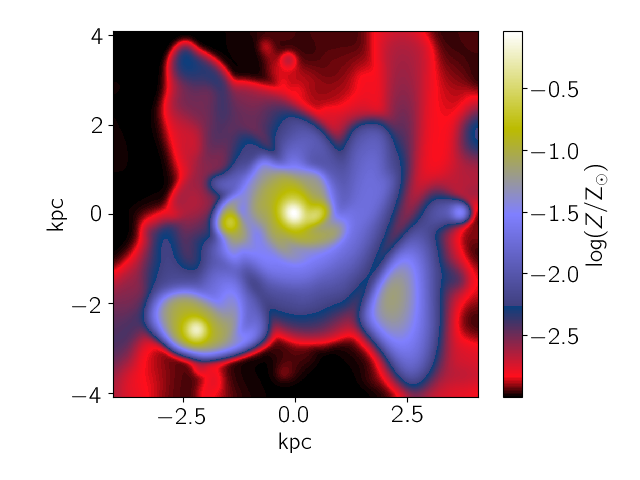}
\caption{
Deviation from the SFR-\CII~and KS relations in Freesia.
We plot the \CII~surface brightness ($\SCII$, {\bf upper left}), SFR surface density ($\dot{\Sigma}_{\star}$, {\bf upper center}), gas surface density ($\Sigma_{g}$, {\bf upper right}), gas metallicity ($Z$, {\bf lower right}), along with the resolved maps for the deviations from the SFR-\CII~ relation (${\rm [CII]}/{\rm [CII]}_{\rm DL}$, eq. \ref{eq_deviation}, {\bf lower left}) and the KS relation ($\kappa_{\rm s}$, eq. \ref{eq_deviation_sk}, {\bf lower center}).
The maps have the same FOV as the one plotted in Fig.s \ref{fig_overview_general} and \ref{fig_map_emission}, but are smoothed with a gaussian kernel with size 100 pc, to allow for a better comparison with observations from \citet{delooze:2014aa}.
\label{fig_delooze_resolved}
}
\end{figure*}

\begin{figure}
\centering
\includegraphics[width=0.485\textwidth]{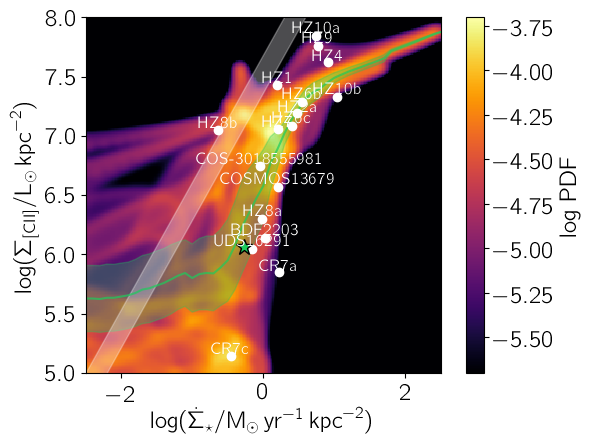}
\caption{
Resolved \CII-SFR relation, i.e. probability density function (PDF) in the $\SCII$-$\dot{\Sigma}_{\star}$ plane of Freesia.
Data for Freesia is taken from Fig. \ref{fig_delooze_resolved}, i.e. SFR and \CII~maps are taken on a 8.2 kpc FOV centred on Freesia and are smoothed over 100 pc. The PDF is weighted uniformly.
The green line (shaded region) marks the average (dispersion) of the distribution in $\sigmasfr$ bins.
The green star is the $\SCII$ and $\dot{\Sigma}_{\star}$ values of the galaxy as probed by a instrument with beam size 2.5 kpc, i.e. similar to the typical ALMA observation.
The \citet{delooze:2014aa} local relation ($\Sigma_{\rm [CII],\ DL}/\surfl$, eq. \ref{eq_cii_observed}) is shown with a grey transparent band, whose thickness marks the dispersions (0.32).
We overplot with white circles the observations of individual components of high-redshift galaxies as analysed in \citet[][in particular see Tab. 2 and Fig. 10 therein]{carniani:2018}, with original data taken from \citet{ouchi2013,capak:2015arxiv,willott:2015arxiv15,maiolino:2015arxiv,jones+17,pentericci:2016apj}.
\label{fig_delooze_resolved_pdf}
}
\end{figure}

\begin{figure}
\centering
\includegraphics[width=0.485\textwidth]{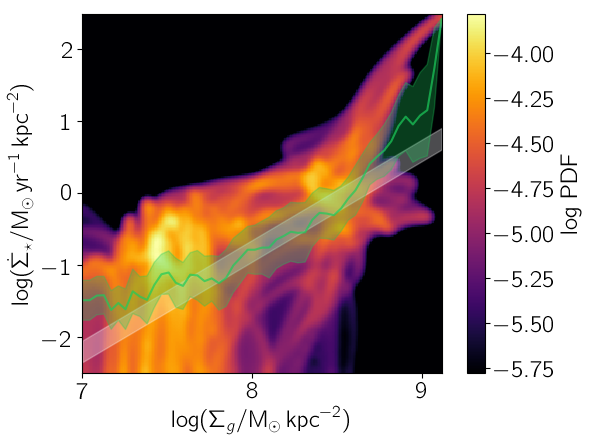}
\caption{
Resolved KS relation, i.e. PDF in the $\dot{\Sigma}_{\star}$-$\Sigma_g$ plane of Freesia.
As for \ref{fig_delooze_resolved_pdf}, data for Freesia is taken from Fig. \ref{fig_delooze_resolved} and the PDF has a uniform weight.
The green line (shaded region) marks the average (dispersion) of the distribution in $\Sigma_g$ bins.
The KS relation ($\dot{\Sigma}_{\star,\,\rm KS}$, eq. \ref{eq_sk_observed}) is plotted with a grey transparent band, whose thickness marks the dispersions (0.15).
\label{fig_sk_resolved_pdf}
}
\end{figure}

In Fig. \ref{fig_cii_sfr} we plot the integrated SFR-\CII~relation for Freesia and the other galaxies that are on average within $(56.9\pm 21.6)\,\kpc$.
For our sample of 11 galaxies we find that \CII~is increasing with SFR, however the slope is shallower with respect to the local \citet{delooze:2014aa} relation: galaxies with lower star formation (${\rm SFR}\lsim 1 \msunyr$) lie above or on top the local relation, while as we go to progressively high rates (${\rm SFR}\gsim 5 \msunyr$) galaxies fall below the relation. The trend for the simulated galaxies is not clear, as the dispersion is large and the sample is limited.

As noted in \citet{carniani:2018}, this trend is similar to what is observed at \highz: it is unclear whether the local \citet{delooze:2014aa} relation holds at high redshift because of the low statistical significance of the observed sample, and there is evidence that the dispersion is larger by a factor $\times 1.8$ with respect to the local \CII-SFR.
In particular, Freesia is within $2\sigma$ from the \citet{delooze:2014aa} relation, similarly to what is observed in high-$z$ galaxies \citep{carniani:2018}.
Low metallicity alone cannot fully explain the tension from the local relation. The mean metallicity of the gas in Freesia and most of the other simulated galaxies is $Z= 0.5\,\zsun$. However, for $Z= 0.5\,\zsun$ the \citet{vallini:2015} model is consistent with the \citet{delooze:2014aa} relation.

Note that at redshift $12\gsim z\gsim 9$ the three simulated galaxies of \citet{katz:2019arXiv} have properties that are consistent with the local \CII-SFR \citep{delooze:2014aa}. However, \citet{katz:2019arXiv} emission lines are computed starting from a library of \code{cloudy} models with constant gas temperature, i.e. not computing the temperature structure of the PDR. The temperature is taken from the hydrodynamical simulation and is computed with a simpler network with respect to \code{cloudy} or other simulations, e.g. not considering non-equilibrium metal cooling \citep[][]{capelo:2018}. The resulting ion abundances and emission lines may be considerably different.

Instead, in the following Sec.s \ref{sec_sk_connection} and \ref{sec_sk_resolved}, we show that falling below the local \CII-SFR relation is connected to the galaxy \quotes{star-burstiness}, i.e. the evolutionary stage in which the galaxy lies above the Kennicutt-Schmidt relation, therefore experiencing an enhanced stellar feedback. The theoretical background for this argument is worked out in detail in a companion work \citep{ferrara:2019}.

\subsection{Connection with Kennicutt-Schmidt relation}\label{sec_sk_connection}

The Kennicutt-Schmidt (KS) relation in our simulated sample is plotted in Fig. \ref{fig_sk_global} (right panel) in the \citet[][]{krumholz:2012apj} \quotes{formulation}, i.e. $\dot{\Sigma}_{\star}$ vs $\Sigma/t_{\rm ff}$. In this formulation the scatter is lower, and the relation holds for single MCs, local and moderate redshift unresolved galaxies.

Freesia and galaxy 1 are under the \citet{delooze:2014aa} relation and above the KS\footnote{Recall that Freesia has recently experienced a burst of star formation (Fig. \ref{fig_time_evol_stars}), this is the reason why it sits above the relation.}, while the opposite holds for galaxies 4 and 9, which are over the local \CII-SFR and below the KS, while galaxies 6 and 8 are consistent with both. This is an indication that the two relations are intimately linked. However galaxies 2, 5, 7, and 10 are below both relations, while 3 is on the \citet{delooze:2014aa} but above the KS.

\subsection{Spatially resolved relations}\label{sec_sk_resolved}

To clarify the connection between KS relation and \CII-SFR relation, we analyse their spatially resolved versions, that are generally considered more fundamental in physical terms than the integrated ones \citep[e.g. for \CII~see][]{herreracamus:2015}.
Freesia is the only galaxy in our simulated sample with a SFR comparable to the one for currently observed galaxies, thus we focus again on that system.

As we are interested in the deviations from the local relations, it is convenient to parameterize them as follows. For the \CII-SFR we use
\begin{subequations}\label{eq_deviation}
\be
{\rm [CII]}/{\rm [CII]}_{\rm DL} \equiv \SCII / \Sigma_{\rm [CII],\ DL}\,,
\ee
where
\be\label{eq_cii_observed}
\log(\Sigma_{\rm [CII],\ DL}) = 1.075 \log(\dot{\Sigma}_{\star}) + 7.51,
\ee
\end{subequations}
with $\SCII$ and $\dot{\Sigma}_{\star}$ in units of ${\surfl}$ and $\surfsfr$, respectively \citep[][data from Tab. 2]{delooze:2014aa}. For the KS we can define
\begin{subequations}\label{eq_deviation_sk}
\be
\kappa_{\rm s} \equiv \dot{\Sigma}_{\star}/\dot{\Sigma}_{\star,\,\rm KS}
\ee
with
\be\label{eq_sk_observed}
\log(\dot{\Sigma}_{\star,\,\rm KS}) = 1.4\log(\Sigma_{g}) -12.0,
\ee
\end{subequations}
where $\Sigma_g$ is in units of $\surfd$ \citep[][see their eq. 2]{heiderman:2010}.

In Fig. \ref{fig_delooze_resolved} we plot the deviation from the relations along with the three quantities determining the local relations ($\SCII$, $\dot{\Sigma}_{\star}$, $\Sigma_g$). To have a fair comparison with observations, the maps are smoothed at 100 pc, similar to the \citet{delooze:2014aa} resolution.

It is striking that in almost all the spots where Freesia is deficient in \CII~($\log ({\rm [CII]}/{\rm [CII]}_{\rm DL} ) < 0$), it is locally star-bursting ($\log \kappa_{\rm s}>0$); this is particularly evident at the centres of both Freesia-A~and Freesia-B, and at the location of the young stars stripped out in the halo of the system.
Vice versa, being above the \citet{delooze:2014aa} relation implies a lack of star formation, as seen in the surroundings of Freesia-B~and in the dense gas filament North-West of Freesia-A.

However, a location above (below) the KS relation is a necessary but not sufficient condition for a gas path to be under (over) luminous in \CII, as seen in the material around the filament and in the outer edges of the system.
This happens because metallicity plays a secondary role in the link between the two relations (Fig. \ref{fig_delooze_resolved}, bottom right). On the one hand, where the metallicity is very low ($Z\lsim 10^{-2}\zsun$), the \CII~is fainter than expected even for high $\Sigma_g$ ($\Sigma_g \gsim 10^8\surfd$), as carbon abundance limits the emission. On the other hand, a low metallicity implies less dust, the main catalyst of \HH~formation, thus consequently stars. This entails a lower $\sigmasfr$ for the patches of gas with similar $\Sigma_g$. {\commento Such low mean metallicity is not found for relatively massive ($M_{\star}\gsim 10^8 \msun$) and star forming (${\rm SFR} \gsim 0.1 \msunyr$) galaxies; low mean ISM metallicities ($Z\lsim 0.05\,\zsun$) are typical of smaller ($M_{\star}\gsim 10^7 \msun$) systems with lower star formation (${\rm SFR} \lsim 0.1\, \msunyr$ \citealt{jeon:2015,jeon:2019}); for these galaxies, metallicity can play a role as relevant as burstiness, however currently no ALMA \CII~detection is available for galaxies with ${\rm SFR} \lsim 1\, \msunyr$.}

The two resolved relations can be further analysed by extracting the PDFs of $\SCII$-$\sigmasfr$ and $\sigmasfr$-$\Sigma_g$. From Fig. \ref{fig_delooze_resolved_pdf} we can see that $\SCII$ is increasing with $\sigmasfr$ in Freesia: the slope is almost flat ($\lsim 0.25$) for low ($\sigmasfr\lsim 0.1 \surfsfr$) and high ($\sigmasfr\gsim 5 \surfsfr$) formation rates, while the trend is nearly linear for intermediate $\sigmasfr$. The scatter in the relation is decreasing with increasing $\sigmasfr$, because of the smaller spread in metallicity of the gas as we go to progressively higher $\SCII$.

The local $\SCII$-$\sigmasfr$ is fitted with an almost linear slope (eq. \ref{eq_cii_observed}) and the observed data range is $10^{5}\lsim \SCII/\surfl \lsim 10^{6.5}$ and $10^{-3}\lsim \sigmasfr/\surfsfr \lsim 1$, with only few observed points (from the galaxy NGC1569) having $10^{-1}\lsim \sigmasfr/\surfsfr \lsim 1$, where the data seem to indicate the presence of a deficit with respect to the linear trend \citep[][see in Fig. 2 therein]{delooze:2014aa}.
The local relation is below the one found for Freesia at $\sigmasfr\lsim 0.1 \surfsfr$, while it is above that at $\sigmasfr\lsim 1 \surfsfr$. The overlap is not perfect in the intermediate region, because of the patch of low surface brightness regions ($\SCII\lsim 10^{5.5}\surfl$) at $0.01\lsim \sigmasfr/\surfsfr\lsim 1$ that drags down the mean relation from Freesia.

In Fig. \ref{fig_delooze_resolved_pdf} we see that most of the $\SCII$-$\sigmasfr$ high-z data -- obtained by integrating the various galaxy components and estimating their UV and IR size \citep[see][for details]{carniani:2018} -- are nicely consistent with the average value extracted from Freesia, while they fall below the local relation. This is an indication those galaxy are dominated by $\sigmasfr\gsim \surfsfr$, critical point where the local $\SCII$-$\sigmasfr$ deviate from linear by saturating to an almost constant value \citep{ferrara:2019}.

To close the loop, we plot the PDF of the KS in Fig. \ref{fig_sk_resolved_pdf}. The local relation and the average found in Freesia are in good agreement up to $\sigmasfr\lsim 10\surfsfr$ ($\Sigma_g\gsim 10^9\surfd$); for higher values, the intense and concentrated star formation and the consequent strong feedback cause these parts of the galaxy to deviate from the averaged KS.
The majority of the regions of the galaxy is below the local $\SCII$-$\sigmasfr$, in particular the patches of gas at high $\sigmasfr$, which are major contributors of its luminosity and SFR; thus, when spatially integrating $\SCII$ and $\sigmasfr$, Freesia results to be below the local \CII-SFR relation and consistent only within $2\sigma$ (Fig. \ref{fig_cii_sfr}).

\section{Conclusions}\label{sec_conclusione}

We have studied the formation and evolution of a sample of Lyman Break galaxies in the Epoch of Reionisation by using crafted, cosmological zoom-in (spatial resolution $\sim 10$ pc) simulations, as part of the \code{SERRA} suite. 

The \code{SERRA} simulations are based on a customised version of the Adaptive Mesh Refinement code \code{ramses} \citep{teyssier:2002}. The ISM thermo-chemical evolution is followed via a non-equilibrium network generated with \code{krome} \citep{grassi:2014mnras}, that allows a precise tracking of the formation of \HH, that can be converted into stars. With respect to previous works \citep{pallottini:2017dahlia,pallottini:2017althaea}, the present simulations perform a full on-the-fly radiative transfer of the  interstellar radiation field (ISRF) thanks to \code{ramses-rt} \citep{rosdahl:2013ramsesrt}.
In the post-processing phase we compute the intensities of several FIR lines (\CII, \NII, and \OIII) from a grid of models obtained with the photo-ionisation code \code{cloudy} V17 \citep{ferland:2017}. These calculations also account for the turbulent structure of the gas \citep{vallini:2015}. This procedure allows a fair comparison of the results with ALMA observations of high-$z$ galaxies.

At $z= 8$, the most massive galaxy in the simulation is \quotes{Freesia}. It has a stellar age of $t_\star \simeq 409\,\myr$, a stellar mass of $M_{\star} \simeq 4.2\times 10^9 \msun$, and a star formation rate of $\SFR\simeq (11.5\pm 1.8)\,\msunyr$, due to a recent burst.
The galaxy is composed by two concentrated ($\sim 200\,\rm pc$) stellar components (\quotes{A} and \quotes{B}) separated by $\simeq 2.5$ kpc. Freesia-A dominates both the mass ($\simeq 85\%$) and star formation ($90\%$) of the system.
Around Freesia, other 11 galaxies are found within $56.9 \pm 21.6$ kpc ($>2$ virial radii); while in our previous work such galaxies were present \citep{pallottini:2017althaea}, their SFR was likely suppressed in their formation stages, due to lack of proper treatment of the ISRF, which was spatially uniform and dominated in intensity by the most massive galaxy.

The properties of Freesia are overall similar to the ones found in previous work at the same evolutionary stage \citep{pallottini:2017althaea}, featuring comparable SFR and $M_\star$. The galaxy develops broken spiral disk of gas with radius $\simeq 0.5\,\kpc$ and average metallicity $Z\simeq 0.5\, \zsun$.
The molecular gas is primarily found at gas densities $n\simeq 200\,\cc$, i.e. about 2$\times$ denser than previously found \citep{pallottini:2017althaea} as a consequence of the higher local ISRF. This is because higher densities are required in order for the gas to self-shield from the stronger Lyman-Werner radiation.
The bulk ($70\%$) of the ionised ISM has low densities ($n = 5 \times 10^{-2} \cc$); $25\%$ is  collisionally ionised by shocks; only $5\%$ is found in the high density regions ($n \gsim 10 \,\cc$), where ionisation is due to the internal presence of young clusters, which gives large fluctuations in the ionising ISRF.

The ISRF of the galaxies features a mean (variance) Habing flux $G = 7.9 G_0$ ($G = 23\, G_0$) and has a rather smooth spatial distribution. The ionisation radiation, parameterized by $U$, shows larger intensity variations in the range $U = 2^{+20}_{-2} \times 10^{-3}$ and a patchy distribution peaked at the location of recent star formation events.
The patchy ionisation structure yields an escape fraction of $f_{\rm esc}\simeq 2\%$, a value consistent with observations of lower redshift galaxies with similar brightness \citep[e.g.][]{bowens:2016,grazian:2017}, and averaged values of simulated galaxies with similar mass \citep[][]{xu:2016apj}. We note, however, that large fluctuations are expected between different evolutionary stages \citep[][]{trebitsch:2017}

The \CII~line luminosity of Freesia is $L_{\rm CII}  = 7.73 \times 10^7\lsun$. The emission extends on a few kpc scales, with peaks around Freesia-A, Freesia-B~and dense non-starforming clumps found near the galaxy. The emission mostly comes from gas with $n \simeq 160\,\cc$ and illuminated by an ISRF intensity $G \simeq 20\,G_0$.
Instead, \OIII~emission is concentrated around Freesia-A, the star-bursting component of the system, and its emission is dominated by gas with a wider range of properties, i.e. $n\simeq (53 \pm 40)\, \cc$ and $U \simeq (6 \pm 2)\times 10^{-3}$. At the Freesia-A~location the oxygen line is very bright, $\SOIII/\SCII\simeq 10$. However, the smaller extent of the \OIII~emitting region implies a lower galaxy-integrated emission, i.e. $L_{\rm OIII} = 2.07 \times 10^7\lsun$.
With respect to \CII, the \OIII~show both a spatial offset ($\sim 2.5 \,\kpc$) and a spectral shift of ($\simeq 120 \kms$), reminiscent of similar evidence found in some systems at high-$z$ \citep{carniani:2017bdf3299}.

Freesia lies below the local \CII-SFR relation \citep[][within $\simeq 2\sigma$]{delooze:2014aa} as it is in a starburst phase, i.e. it sits above the Kennicutt-Schmidt (KS) relation \citep[][within $\simeq 2\sigma$]{schmidt:1959apj,kennicutt:1998apj}.
Spatial analysis reveals that patches of the galaxy that are above the resolved local $\SCII$-$\sigmasfr$ \citet{delooze:2014aa} relation are located below KS, and vice-versa.

In particular, due to the recent starburst, the dense center ($\Sigma_g\gsim 10^9\surfd$) of Freesia lies above the KS ($\sigmasfr\gsim 10\surfsfr$) relation. At such high star formation surface densities the $\SCII$ increases with $\sigmasfr$ less rapidly than expected from the local relation \citet{delooze:2014aa}, which however only covers the low-end of the $\sigmasfr$ range. Thus, the observed \CII-SFR deficit can be primarily ascribed to negative stellar feedback during starburst phases, disrupting molecular clouds around star formation sites. This interpretation was originally proposed in \citet{vallini:2015}; our results fully endorse it. 

Metallicity effects have a weaker impact on the origin of \CII-SFR deviations. Gas with extremely low ($Z\lsim 10^{-2}\zsun$) metallicities fails below both the local $\SCII$-$\sigmasfr$ and the KS relations, because of lack of \CIIion~ions and inefficient~\HH~formation, respectively. However, gas with $Z\lsim 10^{-2}\zsun$ is scarce in Freesia and its combined contribution is subdominant when galaxy integrated relations are considered. This situation is common to most of the galaxies in our sample. {\commento A metallicity effect might be relevant for galaxies that cannot retain their metal production \citep[][]{xu:2016apj,jeon:2019}, that are expected to be smaller ($M_\star \lsim 5\times 10^{6} \msun$) and with a lower star formation (${\rm SFR}\lsim 1 \msunyr$), for which however there is currently no detection with ALMA.} 

\section*{Acknowledgments}

This research was supported by the Munich Institute for Astro- and Particle Physics (MIAPP) of the DFG cluster of excellence \quotes{Origin and Structure of the Universe}.
We thank A. Lupi, M. Trebitsch, J. Rosdahl, and the participants of the \quotes{The Interstellar Medium of High Redshift Galaxies} MIAPP conference for fruitful discussion.
AF and SC acknowledge support from the ERC Advanced Grant INTERSTELLAR H2020/740120.
LV acknowledges funding from the European Union's Horizon 2020 research and innovation program under the Marie Sk\l{}odowska-Curie Grant agreement No. 746119.
We acknowledge use of the Python programming language \citep{VanRossum1991}, Astropy \citep{astropy}, Cython \citep{behnel2010cython}, Matplotlib \citep{Hunter2007}, NumPy \citep{VanDerWalt2011}, \code{Pymses} \citep{Labadens2012}, Pandas \citep{pandasref}, and SciPy \citep{scipyref}.

\bibliographystyle{mnras}
\bibliography{master,codes}
\bsp


\label{lastpage}

\end{document}